\documentclass[useAMS,usenatbib]{mn2e}
\usepackage{url,times,graphicx,amsmath,amsfonts,amssymb,aas_macros,color,epsfig,epstopdf,multirow}

\newcommand{\hMpc}{{\ifmmode{h^{-1}{\rm Mpc}}\else{$h^{-1}$Mpc}\fi}}
\newcommand{\hkpc}{{\ifmmode{h^{-1}{\rm kpc}}\else{$h^{-1}$kpc}\fi}}
\newcommand{\hMsun}{{\ifmmode{h^{-1}{\rm {M_{\odot}}}}\else{$h^{-1}{\rm{M_{\odot}}}$}\fi}}
\newcommand{\ltsima}{$\; \buildrel < \over \sim \;$}
\newcommand{\gtsima}{$\; \buildrel > \over \sim \;$}
\newcommand{\lsim}{\lower.5ex\hbox{\ltsima}}
\newcommand{\gsim}{\lower.5ex\hbox{\gtsima}}
\def\LCDM{$\Lambda$CDM}

\def\lesssim{\mathrel{\hbox{\rlap{\hbox{\lower4pt\hbox{$\sim$}}}\hbox{$<$}}}}
\def\gtrsim{\mathrel{\hbox{\rlap{\hbox{\lower4pt\hbox{$\sim$}}}\hbox{$>$}}}}

\newcommand{\Tab}[1]{Table~\ref{#1}}
\newcommand{\Sec}[1]{Section~\ref{#1}}
\newcommand{\Eq}[1]{Eq.~(\ref{#1})}
\newcommand{\Fig}[1]{Fig.~\ref{#1}}
\newcommand{\beq}{\begin{equation}}
\newcommand{\eeq}{\end{equation}}
\def\beqa{\begin{eqnarray}}
\def\eeqa{\end{eqnarray}}
\def\hMpc{$h^{-1}\,{\rm Mpc}$}
\def\hkpc{$h^{-1}\,{\rm kpc}$}
\def\LCDM{\ensuremath{\Lambda}CDM}

\def\head{
 \vbox to 0pt{\vss
                   \hbox to 0pt{\hskip 440pt\rm LA-UR-10-07069\hss}
                  \vskip 25pt}}

\title[Applying scale-free mass estimators to CLUES]
{Applying scale-free mass estimators to the Local Group in Constrained Local Universe Simulations}
\author[Di Cintio et. al]
       {Arianna Di Cintio$^{1}$\thanks{E-mail: arianna.dicintio@uam.es}, Alexander Knebe$^{1}$, Noam I. Libeskind$^2$,  Yehuda Hoffman$^3$,\newauthor Gustavo Yepes$^1$, Stefan Gottl\"ober$^2$\\
$^{1}$Departamento de F\'isica Te\'orica, M\'odulo C-8, Facultad de Ciencias, Universidad Aut\'onoma de Madrid, 28049 Cantoblanco, Madrid, Spain\\
$^2$Leibniz-Institut $f\ddot{u}r$ Astrophysik Potsdam (AIP), An der Sternwarte 16, D-14482 Potsdam, Germany\\
$^3$Racah Institute of Physics, The Hebrew University of Jerusalem, 91904 Givat Ram, Israel
}

\begin{document}

\date{Accepted XXXX . Received XXXX; in original form XXXX}

\pagerange{\pageref{firstpage}--\pageref{lastpage}} \pubyear{2010}

\maketitle

\label{firstpage}


\begin{abstract}

We use the recently proposed scale-free mass estimators to determine the masses of the Milky Way (MW) and Andromeda (M31) galaxy in a dark matter only Constrained Local UniversE Simulation (CLUES).  While these mass estimators work rather well for isolated spherical host systems, we examine here their applicability to a simulated binary system with a unique satellite population similar to the observed satellites of MW and M31. We confirm that the scale-free estimators work also very well in our simulated Local Group galaxies with the right number of satellites which follow  the observed radial distribution. In the isotropic case and under the assumption that the satellites are tracking the total gravitating mass, the power-law index of the radial satellite distribution $N(<r)\propto r^{3-\gamma}$ is directly related to the host's mass profile $M(<r)\propto r^{1-\alpha}$ as $\alpha=\gamma-2$.  The use of this relation for any given $\gamma$ leads to highly accurate mass estimations which is  a crucial point for observer, since they do not know a priori the mass profile of the MW and M31 haloes. We discuss possible bias in the mass estimators and conclude that the scale-free mass estimators can  be satisfactorily applied to the real MW and M31 system.

\end{abstract}
\noindent
\begin{keywords}
  methods: $N$-body simulations -- galaxies: haloes -- galaxies:
  evolution -- cosmology: theory -- dark matter
\end{keywords}

\section{Introduction} \label{sec:introduction}
Although measurements of gas rotation curves are often precise enough to constrain the inner most mass of galaxies like the Milky Way (MW) and Andromeda (M31) (within a few tens of kpc), kinematics of a tracer populations are needed to compute the mass within greater radii. These tracers can either be globular clusters or planetary nebulae (e.g. \citet{Schuberth10};\citet{Woodley10}), halo stars \citep{Xue08} or satellite galaxies (e.g. \citet{Watkins10}).
Since the kinematics of these objects are determined by the underlying host potential they allow for an estimate of the enclosed mass within their respective distances from the center of the host. 

Kinematic data of galaxies in clusters have already been used to compute the mass profiles and galaxy orbits in nearby clusters \citep{Wojtak10}; moreover, the mass of four Milky Way dwarf spheroidals (dSphs) satellites were constrained with high precision thanks to kinematic data sets \citep{Lokas09}. Line-of-sight kinematic observations enable accurate mass determinations at half-light radius for spherical galaxies such as the MW dSphs \citep{Wolf10}: at both larger and smaller radii however, the mass estimation remains uncertain because of the unknown velocity anisotropy.

Regarding our own Galaxy, having position and proper motion data of the MW's satellite galaxies would allow one to satisfactorily apply the great majority of kinematic mass estimators to the calculus of the Milky Way's mass, including the recently proposed  ``scale-free projected mass estimator" \citep[hereafter W10]{Watkins10}.

In the very near future the knowledge of the full six-dimensional phase-space information for all objects, in the close Universe, brighter than $G\approx20$ mag, is going to be dramatically improved thanks to space missions, like GAIA\footnote{\texttt{http://www.gaia.esa.int}}, whose goal is to create the largest and most precise three dimensional chart of the Milky Way by providing precise astrometric data like positions, parallaxes, proper motions and radial velocity measurements for about one billion stars in our Galaxy and throughout the Local Group (LG). 

\citet{An11} recently showed that new proper motions data with the targeted GAIA accuracies will be able to outperform the presently existing line-of-sight based mass estimators.
But until the proper motions of these satellite galaxies become available, one needs to rely on assumptions and simplifications.

One of the first estimators of the mass contained within the LG is based on the ``timing" argument of \citet{Kahn59}.
More accurate mass estimators for spherical systems are based either on the virial theorem or on the moments of projected mass, as first introduced by \citet{Bahcall81}. 
They assumed that only projected distances and line-of-sight velocity information were available, and demonstrated the goodness of the projected mass estimator. The main advantages of such a projected mass estimator over the virial theorem, neglecting the uncertainties in the eccentricity distribution, are that they are unbiased, their variance is known, and they converge to the real mass with an error proportional to $N^{-1/2}$, where $N$ is the sample number. Moreover, the information from every tracer particle is equally weighted, contrary to what happens for the virial theorem case.

Previous studies successfully used these mass estimator methods to ``weigh'' M31; and more recently, W10 developed alternative forms of estimators that can also be applied to the calculus of the MW's mass: they rely on the assumption that both the host galaxy and its distribution of tracer objects are spherically symmetric. What is still unclear however, is the shape of the MW and M31 halo,  with various authors in the literature disagreeing over whether its triaxial \citep{Law09} or spherical \citep{Koposov10}.

\citet{Deason11} and \citet{Evans11} have demonstrated the statistical validity of the W10 mass estimators using a set of 431 parent haloes and 4864 associated satellite galaxies, taken from the GIMIC simulations \citep{Crain09}: under the assumption of having a host profile of the type NFW \citep{Navarro96}, they found that the fraction of estimated halo mass which lies within a factor of two of the true mass is about $80\%$.

In this work we aim to gauge the quality of the method introduced in W10 by using the Local Group identified in the WMAP5 dark matter only constrained cosmological simulation of the CLUES project\footnote{\texttt{http://www.clues-project.org}}, a numerical laboratory for testing the applicability of such a method to the MW and M31.

Observational data of the nearby Universe are used to constrain the initial conditions of the CLUES simulations. 
These constrained simulations, in which the Local Group lies in the right cosmological environment, provide a complementary approach, with respect to cosmological simulations, to make a comparison between numerical results and observations. Thus, verifying the robustness of the W10 mass estimators in our unique simulated LG is an important test in addition to the more statistical methods offered by cosmological simulations \citep{Deason11}.

The idea is to verify whether these estimators can accurately be applied to a system such as the one found in our LG and composed of the Milky Way and the Andromeda galaxy.
The arrangement and formation history of this galactic binary system, according to our present state-of-the-art of numerical simulations, is rather unique and involves preferential infall directions of their subhaloes \citep{Libeskind10infall}, a backsplash population \citep{Knebe11a}, and even renegade satellites \citep{Knebe11b}, i.e. satellites that change their affiliation from one of the two hosts to the other. Furthermore, the MW and M31 satellites do in fact remember the non-random nature of their infall after several orbits \citep{Libeskind12}. We also need to mention that -- when comparing constrained against un-constrained simulations -- only 1-3\% of the Local Group candidates share similar formation properties \citep{Forero11}.

Moreover, the observed Milky Way satellites are found to be highly anisotropical, lying within a thin disc which is inclined with respect to the MW's one, with a minor-to-major axis ratio $c/a\approx0.3$: this flattened distribution is not compatible with the satellites to have been randomly selected from an isotropic subset \citep[]{Kroupa05,Metz07,Metz08}.
Previous cosmological simulations showed anisotropy in the subhaloes population, with the brightest satellites distributed along disk-like structures, consistently with the observed MW satellites \citep[cf.][]{Knebe04,Libeskind05,Zentner05}.
This anisotropy, which is also observed in our simulated subhaloes, may in principle cause a bias in the application of the mass estimator, since the hypothesis of spherical symmetry is broken.  

\textit{We therefore raise (and answer) the question about the applicability of scale-free mass estimators to such a special system as the Local Group.}

\section{The Simulation} \label{sec:simulation}
Here we briefly describe the simulation and the way (sub-)haloes have been identified within it.

\subsection{Constrained Simulations of the Local Group}
\label{sec:localgroup}
The dark matter only simulation used here forms part of the Constrained Local UniversE Simulations (CLUES) project  and is designed to reproduce the Local Group of galaxies within a WMAP5 cosmology \citep{Komatsu09}, i.e. $\Omega_{m} = 0.279$, $\Omega_{b} = 0.046$, $\Omega_{\Lambda} = 0.721$. We use a normalization of $\sigma_8 = 0.817$ and a slope of the power spectrum of $n=0.96$. We used the TreePM $N$-body code \texttt{GADGET2} \citep{Springel05} to simulate the evolution of a cosmological box with side length of $L_{\rm box}=64 h^{-1} \rm Mpc$. Within this box we identified (in a lower-resolution run utilizing $1024^3$ particles) the position of a model local group that closely resembles the real Local Group \citep[cf.][]{Libeskind10}. This Local Group has then been re-sampled with 64 times higher mass resolution in a region of $2 h^{-1} \rm Mpc$ about its centre giving a nominal resolution equivalent to $4096^3$ particles giving a mass resolution of $m_{\rm   DM}=2.95\times 10^{5}$\hMsun.
Within this environment we identified two main haloes, formally corresponding to the Milky Way and the Andromeda galaxy, whose main properties are listed in \Tab{tab:MWM31}, together with their corresponding actual observational properties. The virial mass of each halo is in units of $10^{12}M_{\odot}$, while the virial radius and the distance between the two hosts, listed as $D$, are in Mpc. Both these quantities are based upon the definition $M(<R_{\rm vir})/(4\pi/3R_{\rm vir}^3)=\Delta_{\rm vir} \rho_b$ where $\rho_b$ is the cosmological background density and $\Delta_{\rm vir}=354$ for the considered cosmology and redshift $z=0$. The concentration is $c_2=R_{\rm vir}/r_2$, where $r_2$ denotes the "scale radius" where the product $\rho(r)r^2$ reaches its maximum value. The two axis ratios $b/a$ and $c/a$ are derived from the eigenvalues $a>b>c$ of the moment of inertia tensor, and the vertical-to-planar axis ratio is reported for M31. The $\alpha$ parameter is the exponent corresponding to a scale-free host mass profile $M(r)\propto r^{1-\alpha}$, see \Sec{sec:formulae} for more details. The observationally derived masses are based on the work of W10, and represent the estimates of each galaxy mass assuming a virial radius of 300 kpc, using the observed anisotropy parameter $\beta$ and including satellites' proper motions. 

\subsection{The (Sub-)Halo Finding}
\label{sec:halofinding}
In order to identify haloes and subhaloes in our simulation we have run the MPI+OpenMP hybrid halo finder \texttt{AHF}\footnote{\texttt{AMIGA} halo finder, to be downloaded freely from \texttt{http://www.popia.ft.uam.es/AMIGA}} described in detail in \cite{Knollmann09}. \texttt{AHF} is an improvement of the \texttt{MHF} halo finder \citep{Gill04a}, which locates local overdensities in an adaptively smoothed density field as prospective halo centers. We would like to stress that our halo finding algorithm automatically identifies haloes, sub-haloes, sub-subhaloes, etc. and it can reliably recover substructures containing at least 30 particles \citep{Knebe11c}. For more details on the mode of operation and actual functionality we though refer the reader to the code description paper by \citet{Knollmann09}, while an in-depth comparison to other halo finding techniques can be found in \citet{Knebe11c} and Onions et al., in preparation.
A complete summary of the characteristic of the subhaloes population of the two main haloes, MW and M31, is shown in \Tab{tab:MWM31}, together with a comparison of the properties of their observed satellite galaxies. The $r_{\rm out}$ and $r_{\rm in}$ are the radius of the outermost and innermost tracer, respectively, in Mpc (in the case of M31 we listed the projected distances). The quantity $N_{\rm sat}$ represents the number of simulated subhaloes (or observed satellite galaxies) within $0.3$~Mpc from each host center. 

\begin{table}
 \caption{Main properties of the two haloes (representing the MW and M31 galaxy, respectively) considered in this work, and of their respective subhaloes population. The virial mass of each halo is in units of $10^{12}M_{\odot}$, while the virial radius and the distance between the two hosts, $D$, are in Mpc. We listed the observational inferred quantities of MW and M31, that refer to the work of: (a) W10, (b) \citet{Law09}, (c) \citet{Ban08}, (d) \citet{McConnachie05}, (e) \citet{Mateo98}, (f) \citet{Ibata07}, (g) \citet{Martin08}, (h) \citet{Kara04}.}
\begin{center}
\begin{tabular}{lcccc}
\hline
\hline
property                                & MW        &  M31       & MW        &  M31       \\
                                    & \multicolumn{2}{c}{simulation} & \multicolumn{2}{c}{observed}\\
\hline
$M_{\rm vir} $                                    &$ 1.674$    & $ 2.226$   & $2.7\pm0.5^{(a)} $&$ 1.5\pm0.4^{(a)}$\\
$R_{\rm vir}$                                      & 0.310    &  0.340      & $0.300^{(a)}$ & $0.300^{(a)}$ \\
$c_2$                                                      & 11.7     &  10.7  & - & - \\
$b/a$  & 0.937   & 0.978   & $0.83^{(b)}$   & \multirow{2}{*} {$0.4^{(c)}$} \\
$c/a$   & 0.883    & 0.872   & $0.67^{(b)}$  \\
$\alpha$                                              & -0.034 &-0.052 & -&-  \\
$D$			      &\multicolumn{2}{c}{0.782}  &\multicolumn{2}{c}{$0.785\pm0.025^{(d)}$} \\
\hline
$N_{\rm sat}$                       & 1205     & 1405     & 24 & 21 \\
$r_{\rm out}$			               &0.309     & 0.340   &$0.250\pm0.003^{(e)}$ &    $0.270^{(f)}$  \\
$r_{\rm min}$           			      & 0.018     &0.014    & $0.023\pm0.002^{(g)}$ &  $0.005^{(h)}$ \\
\hline 
\hline
\end{tabular}
\end{center}
\label{tab:MWM31}
\end{table}

\section{Scale-free mass estimators} \label{sec:theory}
Even though the mass estimators are derived under the assumption that the respective distributions are scale-free, they have nevertheless been successfully applied to the observed MW and M31 (W10) where the hierarchical structure formation model supports the notion that the density profile of dark matter haloes follows the functional form originally proposed by \citet{Navarro96}, i.e. the so-called NFW profile.

\citet{Xue08} constrained the mass distribution of the MW's dark matter halo by analyzing the kinematic of thousands of blue horizontal-branch halo stars, finding a profile that is consistent with a combination of a fixed disk and bulge model with a NFW dark matter halo.
\citet{Seigar08} have derived new mass models for M31, and found that while a NFW and an adiabatically contracted NFW profiles can both produce reasonable fits to the observed rotation curve of M31, the pure NFW model requires a halo concentration too high with respect to the range predicted by the \LCDM\ cosmology, and is therefore disfavoured.
Thus, it is still debatable whether the Milky Way and Andromeda galaxy haloes actually follow a NFW profile.

In this Section we briefly introduce the scale-free mass estimators, which are directly taken from W10: we refer the reader to their work for a derivation of the respective formulae.

\subsection{Theory of mass estimators}\label{sec:formulae}
Here we present the four relevant formulae and the three parameters each formula depends on: we see that the mass estimator takes different forms according to the available informations from the tracer populations.

\subsubsection{Full Information Estimator (FIE)}
 In the optimum case that the full six-dimensional phase-space information is accessible, the mass estimator can be written as:

\beq
M(<r_{\rm out})= \frac{C}{G}\frac{1}{N} \sum_{i=1}^{N_{\rm tracer}}v_i^2r_i^\alpha,\label{full_info}
\eeq
\beq
\textrm{with} \hspace{0.25cm}C= \frac{\alpha+\gamma-2\beta}{3-2\beta}r_{\rm out}^{1-\alpha}           
\eeq

\noindent
Where $v$ and $r$ are the velocity and distance of each individual tracer particle, $r_{\rm out}$ represents the radius of the outermost tracer, and $G$ is the gravitational constant. The dimensional constant $C$ is constructed out of three additional parameters determined by the host potential ($\alpha$), the tracer's radial distribution ($\gamma$), and the tracer's velocity anisotropy ($\beta$), more details in \Sec{sec:parameters_descr} where these parameters are algebraically defined. Note that we can only estimate the halo mass contained within the outer radius $r_{\rm out}$ set by the distance to the farthest tracer. The mass is then constructed as an average of $v^2r^\alpha$ over the total number of tracer objects, $N_{\rm tracer}$. We will refer to \Eq{full_info} as the Full Information Estimator or simply FIE.

\subsubsection{Radial Information Estimator (RIE)}
In the case that only the radial velocity, with respect to the center of the host galaxy, and the individual distances of the tracer population are known, $v_r$ and $r$ respectively, a different definition of the constant $C$ must be used:

\beq
M(<r_{\rm out})= \frac{C}{G}\frac{1}{N}  \sum_{i=1}^{N_{\rm tracer}} v_{r,i}^2r_i^\alpha,\label{radial_info} 
\eeq

\beq
\textrm{with}\hspace{0.25cm}C= (\alpha+\gamma-2\beta)r_{\rm out}^{1-\alpha}          
\eeq

\noindent
We shall call this the Radial Information Estimator, RIE: this case applies to our own Milky Way. Since we do not have the proper motion of all of its satellites, but just of 9 of them (see for instance \citet{Metz08}), it is safer to assume the RIE.
It must be noticed that in absence of proper motion $v_r$ may be calculated from $v_{los}$ by using the statistical correction:

\beq
<v_r^2>=\frac{<v_{los}^2>}{1-\beta \sin^2\phi}\label{correction}
\eeq

\noindent
where $\phi$ is the angle between the vector from the galactic centre to the satellite and the vector from the sun to the satellite. As we can see, this correction further depends from the anisotropy parameter $\beta$. We will come back to the proper placement of the observer and the relevance of this correction, respectively, later on.

\subsubsection{Line-of-Sight Information Estimator (LIE)}
When using only projected line-of-sight velocities $v_{los}$ and actual distances $r$ for the tracer population, the mass estimator referred to as the Line-of-sight Information Estimator, or LIE, and may be written as:

\beq
M(<r_{\rm out})= \frac{C}{G}\frac{1}{N}  \sum_{i=1}^{N_{\rm tracer}} v_{los,i}^2r_i^\alpha,\label{los_info}  
\eeq

\beq
\textrm{with}\hspace{0.20cm} C= \frac{3(\alpha+\gamma-2\beta)}{3-2\beta}r_{\rm out}^{1-\alpha}         
\eeq

\noindent
This estimator must be used, for example, when calculating the mass of the Andromeda galaxy.

\subsubsection{Projected Information Estimator (PIE)}
In the worst case scenario in which the only data available are both projected distances $R$ and line-of-sight velocities $v_{los}$ for the tracer population, the corresponding estimator is:

\beq
M(<r_{\rm out})= \frac{C}{G} \frac{1}{N} \sum_{i=1}^{N_{\rm tracer}} v_{los,i}^2R_i^\alpha,  \label{projected_info}       
\eeq

\beq
\textrm{with}\hspace{0.2cm}C= \frac{\alpha+\gamma-2\beta}{I_{\alpha,\beta}}r_{\rm out}^{1-\alpha}  
\eeq
\noindent
where 

\beq
I_{\alpha,\beta}=\frac{\pi^{1/2}\Gamma(\frac{\alpha}{2}+1)}{4\Gamma(\frac{\alpha}{2}+\frac{5}{2})}\left[\alpha+3-\beta(\alpha+2)\right]
\label{i_capital}
\eeq

\noindent
and $\Gamma(x)$ is the gamma function. We will refer to this last equation as the Projected Information Estimator, PIE.

\subsubsection{The parameters $\alpha$, $\beta$, and $\gamma$}\label{sec:parameters_descr}
The ever present constant $C$ is composed of three parameters, describing the host potential as well as particulars of the tracer population, under the assumption that they both can be sufficiently described by scale-free models. We further assume spherical symmetry for our tracer population.

The $\alpha$ parameter corresponds to a scale-free gravity field, which is equivalent to a host mass profile of the form:

\beq
M(r)\propto r^{1-\alpha} \label{mass_profile}
\eeq

\noindent 
or, equivalently, to a mass density that scales as $\rho \propto r^{-(\alpha+2)}$.\\

The $\beta$ parameter is the Binney velocity anisotropy parameter \citep{Binney87}, defined as:
 
\beq
 \beta=1-\frac{\sigma_t^2}{2\sigma_r^2} \label{beta}
\eeq

\noindent
in which $\sigma_t^2$ and $\sigma_r^2$ are the tangential and the radial velocity dispersions of the tracer objects. $\beta$ provides information about the orbital distribution of our tracer population. \\

Lastly, the $\gamma$ parameter represents the exponent of the power law describing the radial number density distribution $n(r)$ of the tracer population:

\beq n(r)\propto r^{-\gamma}. \label{gamma}
\eeq

These three parameters are fundamental in describing the geometry of the system and, together with the kinematical information of the tracers, allow us to compute an accurate estimator of the mass of a host halo. It is thus absolutely essential that they are determined with the highest possible accuracy. 
In reality however, this is not always possible: we are often forced to make assumptions regarding the form of the underlying host potential. Moreover, the number of the known satellites of both Milky Way and Andromeda is only $\sim25$, making the determination of the $\gamma$ parameter relatively inaccurate. In addition, in the Milky Way's case, only 7 of these objects have accurately measured proper motions: with such a small sample the velocity anisotropy $\beta$ is widely unconstrained by data.
In the next \Sec{sec:massdependance} we will present the dependence of the mass estimator on each of these parameters, computing -- for a specific case -- the error introduced by uncertainty in $\alpha$, $\beta$, and $\gamma$, respectively.

\subsection{Dependence of the mass estimators on the parameters} \label{sec:massdependance}

Given the inherent  inaccuracy in determining the three model parameters, we would like to gauge the sensitivity of the mass estimators to their uncertainties, considering an adequate set of subhaloes covering a radial range out to $r_{\rm out}$.

\subsubsection{$\gamma$ dependence}

We aim to study the dependence of the mass estimator on the parameter $\gamma$, which represents the exponent describing the radial distribution of the satellites population. We therefore calculated the relative variation of the estimated mass per variation in $\gamma$:

\beq
\frac{\Delta M}{M \Delta\gamma}=\frac{1}{\alpha+\gamma-2\beta} \label{mass_relative_variation}
\eeq

\noindent
We note that \Eq{mass_relative_variation} is valid for the four cases of FIE, RIE, LIE and PIE, being independent from the radial distribution of satellites; it provides a tool for calculating the expected uncertainty in the mass determination given the expected errors in  $\gamma$.

Assuming an isotropic distribution of orbits, i.e. $\beta=0$, we focus on the real case scenario in which the $\gamma$ parameter is $\gamma\cong2$ (as found in W10 for the observed satellites of MW and M31): allowing for an uncertainty of $\Delta\gamma/\gamma\sim25\%$, and recalling the typical value for $\alpha$ that is around 0.55 for a NFW host (e.g., W10), we see that the error in the estimated mass is as high as $\Delta M/M\sim20\%$. 
This error will be even larger when considering smaller value of $\gamma$ and $\alpha$, as well as for $\beta>0$.

\subsubsection{$\beta$ dependence}\label{beta_dependence}
\begin{figure*}\begin{center}
$\begin{array}{cc}  \includegraphics[width=3in]{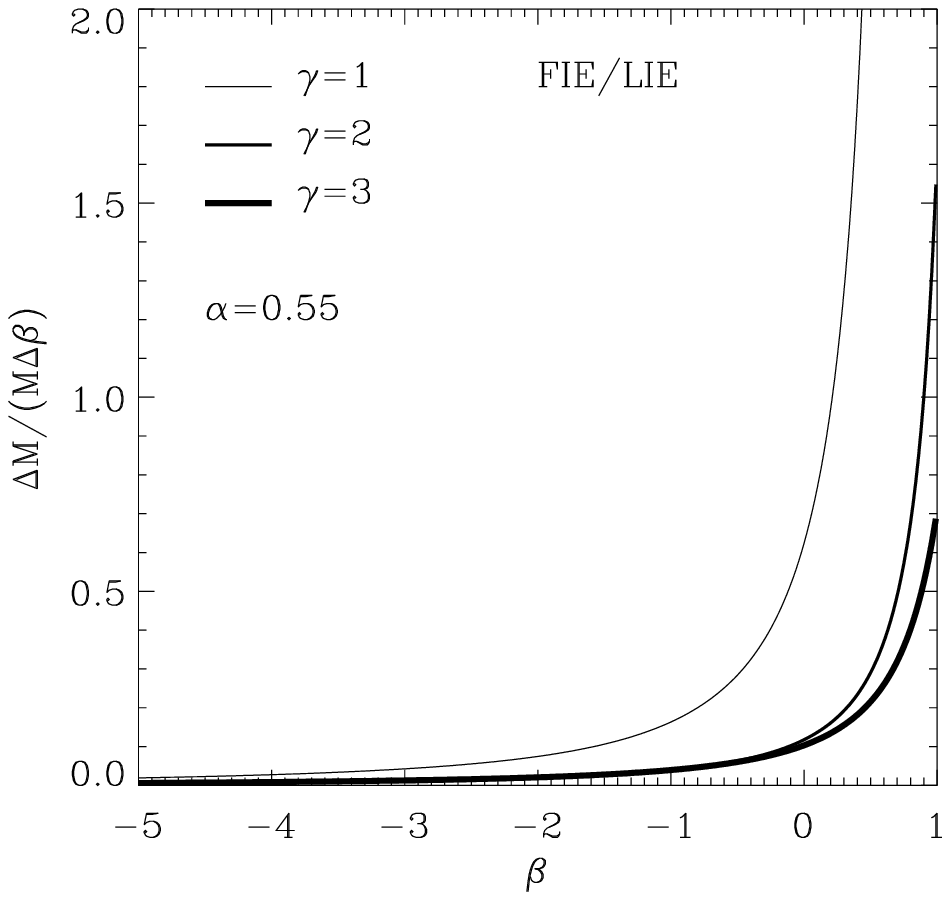}\includegraphics[width=3in]{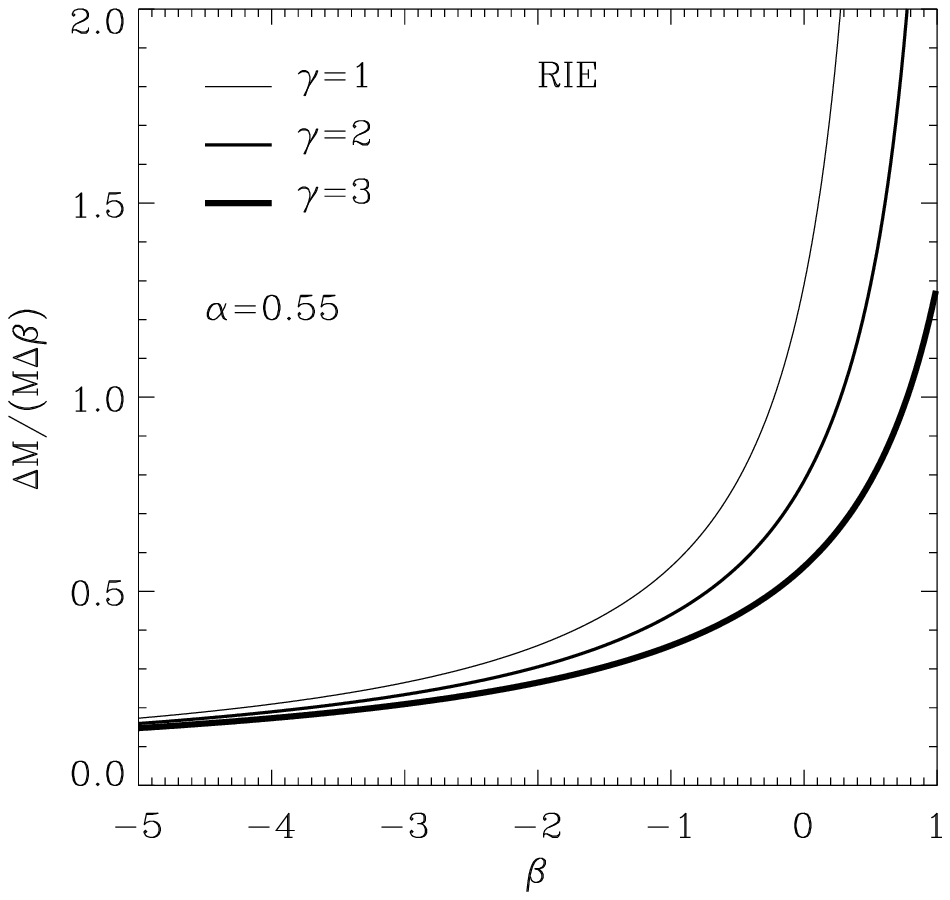}\end{array}$
  \caption{Relative variation per unit parameter change of the mass estimated as a function of the $\beta$ parameter in the case of FIE/LIE (left panel) and RIE (right panel) estimators. The fractional error in the mass estimation is larger for the RIE estimator, used for MW, than for the FIE/LIE estimators, applicable to M31. See full text for more details.}
\label{fig:beta_variation}
\end{center}
\end{figure*}
Regarding the changes in the mass estimation due to the anisotropy parameter, we recall that this parameter is directly obtained from the velocities of the tracer population, computing the tangential and the radial velocity dispersion of each subhalo (as opposed to the $\alpha$ and $\gamma$ parameter, which are derived by assuming a power law distribution).
The $\beta$ average value has been found to be $\beta\sim-0.3$ and $\beta\sim-0.02$ in our CLUES simulation, for the full set of subhaloes of the MW and M31, respectively.
While these values of $\beta$ indicate that we are close to the isotropic regime, i.e. $\beta=0$, the MW anisotropy parameter slightly favors tangential orbits, in agreement with the measured proper motions of the known MW satellites. 
It is thus essential to understand how the variation in the $\beta$ parameter affects the determination of the host mass. For the RIE estimator the corresponding equation reads 

\beq
\left(\frac{\Delta M}{M\Delta \beta}\right)_{RIE}=\frac{-2}{\alpha+\gamma-2\beta} \label{mass_relative_variation_beta_RIE}
\eeq

\noindent
whereas for the FIE and LIE cases it is

\beq
\left(\frac{\Delta M}{M\Delta \beta}\right)_{FIE,LIE}=\frac{2}{3-2\beta}-\frac{2}{\alpha+\gamma-2\beta}\label{mass_relative_variation_beta_FIE_LIE}
\eeq

\noindent
finally, for the PIE scenario

\beq
\left(\frac{\Delta M}{M\Delta \beta}\right)_{PIE}=\frac{\alpha+2}{\alpha+3-\beta(\alpha+2)}-\frac{2}{\alpha+\gamma-2\beta}.\label{mass_relative_variation_beta_PIE}
\eeq

\noindent
In \Fig{fig:beta_variation} we present the absolute value of the fractional mass variation as a function of the $\beta$ value for the FIE and LIE cases (left panel) and for the RIE estimator (right panel). We do not plot the mass changes in the PIE case, as it is practically identical to the FIE and LIE ones.
As in the previous section, considering the general case of having a NFW halo, with values of $\beta$ close to zero and $\gamma=2$, which is the usual case for the hosts considered here and elsewhere (e.g., W10), we find that the error due to variations of $\Delta\beta=\pm1$ for the FIE, LIE and PIE estimators is actually quite low and is below 10\% for $\alpha=0.55$. 
Moreover, \citet{Evans11} found that for much of the radial regime covered by the tracer population, any variation of the anisotropy parameter within its physical range leads to the same estimator in the case of the PIE scenario. Thus, in the case of an external galaxy whose dark matter halo follows a NFW profile with $\alpha=0.55$ and $2<\gamma<3$, we can assume to have a minor error due to $\beta$: the major uncertainty in the mass estimation comes from the assumption made on the $\alpha$ and $\gamma$ parameters. This last statement is valid for the FIE, LIE and the PIE estimator: it does not matter if we have real satellites distances or projected ones, the biggest error on the mass does not come from the anisotropy parameter.\\

The situation is however, completely different for the MW galaxy, for which the RIE formula holds, i.e. we have radial information on the satellite velocities. In this case, a variation of $\Delta\beta=\pm1$ could cause an error in the mass estimation of around $80\%$ if we have $\gamma=2$ and $\alpha=0.55$. Therefore, the $\beta$ parameter is unfortunately the greatest concern in the calculus of the mass of our own Galaxy. Please note that if we knew the three dimensional velocities of the MW satellites as opposed to only the radial ones, we would be dealing with \Eq{mass_relative_variation_beta_FIE_LIE}, thus being in the regime in which the correct evaluation of the $\beta$ parameter will only have a subordinate influence.

We close by remarking that this discussion perfectly agrees with the previous study of the influence of the $\beta$ parameter on the mass estimation as presented in W10.

\subsubsection{$\alpha$ dependence}

Finally, we computed the amount or error introduced by uncertainties in the $\alpha$ parameter, which is directly connected to the potential of the host halo.
The fractional variation of the estimated mass, for the FIE estimator, is:

\beq
\left(\frac{\Delta M}{M \Delta\alpha}\right)_{FIE}=\frac{1}{\alpha+\gamma-2\beta} - \ln(r_{\rm out}) + \frac{\sum_{i}  v^2_ir^{\alpha}_i \ln(r_i)}{\sum_{i}  v^2_ir^{\alpha}_i}  \label{mass_relative_variation_FIE}
\eeq

\noindent
where the summation, as usual, is performed over the total number of tracers $N_{\rm tracer}$ and $r_{\rm out}$ is the radius of the outermost subhalo.
\Eq{mass_relative_variation_FIE} is formally identical for the RIE and LIE case as well, after substituting the full velocity $v$ with the radial velocity $v_r$ or the line-of-sight $v_{los}$ one, respectively.

When dealing with the PIE scenario, instead, the error can be calculated through the following equation :

\begin{eqnarray}
\left(\frac{\Delta M}{M \Delta\alpha}\right)_{PIE}=\frac{1}{\alpha+\gamma-2\beta} - \ln(r_{\rm out}) + \frac{\sum_i  v^2_{los,i}R^{\alpha}_i \ln(R_i)}{\sum_i  v^2_{los,i}R^{\alpha}_i} +
\nonumber\\
  +\frac{\Psi(\frac{\alpha}{2} + \frac{5}{2})}{2}  - \frac{\Psi(\frac{\alpha}{2} + 1)}{2} - \frac{1-\beta}{\alpha +3-\beta(\alpha+2)}           \label{mass_relative_variation_PIE}
\end{eqnarray}

\noindent
where $\Psi(x)$ is the digamma function, defined as the derivative of the logarithm of the $\Gamma(x)$ function.

Unlike the other cases, we can not give a generalized estimation of the error introduced by the $\alpha$ parameter, it being dependent on the radial distribution of the satellites population: this uncertainty varies for every specific scenario and needs to be calculated individually. 

\section{Application to the CLUES simulation}\label{sec:appl_to_clues}
We now move to the application of the scale-free mass estimators to a situation as close as possible to our Local Group. To this extent we use the CLUES simulation introduced in \Sec{sec:simulation}. While we are certain that the scale-free approximation leads to credible results as shown by W10, \citet{Evans11} and \citet{Deason11}, it remains to be seen whether the uniqueness of the Local Group with its binary host system and particular formation history involving preferential infall \citep{Libeskind10infall}, renegade satellites \citep{Knebe11b} and anisotropically distributed subhaloes (not explicitly shown here) will effect the mass estimate.  Moreover, we would like to gauge the accuracy of these mass estimators when fewer tracers  are used, as in the real LG.

From now on we will refer to the case of an observer that is placed at the center of our Galaxy and looking towards the MW's satellites or to the nearby M31's ones. The choice to put the observer in the galactic center instead that at the solar radius may affect the determination of $v_{r}$ from $v_{los}$, given the fact that the radial velocity should be computed with respect to the sun. However, this is practically identical to the radial velocity with respect to the galactic center for distant tracers, for which $\sin \phi \sim 0$, which is the case for our subhaloes. Moreover, the anisotropy parameter $\beta$, which appears in the correction factor of \Eq{correction}, has always been found to be very close to zero in our simulations (as reported in \Sec{beta_dependence}). Nevertheless, when applying the different mass estimators we also used the correction factor given by  \Eq{correction}, placing the observer on a sphere of radius $8$~kpc from the galactic center, and we verified that the affect of this correction is at the $<0.5\%$ level. We will thus refer, through the paper, to the case of an observer placed in the galactic center.

\subsection{Obtaining the parameters $\alpha$, $\beta$ and $\gamma$} \label{sec:parameters}
In order to apply the mass estimator method to our simulated galaxies, we  need to calculate the three unknowns $\alpha,\beta$ and $\gamma$ that appear in \Eq{full_info},\eqref{radial_info},\eqref{los_info} and \Eq{projected_info}. 

\subsubsection{The satellite parameters: $\gamma$ and $\beta$} \label{sec:beta_gamma}

The $\gamma$ parameter is simply obtained by fitting the radial number distribution $N(<r)$ of each host's subhaloes to the functional form

\beq
N(<r)\propto r^{3-\gamma} \label{number_gamma},
\eeq

\noindent
assuming that the number density $n(r)$ follows \Eq{gamma}.\\

The velocity anisotropy parameter $\beta$, as defined in \Eq{beta}, is obtained by first calculating the radial velocity dispersion of the subhaloes, projecting their velocities along the radial axis, then by computing the tangential component of $\sigma$ through the relation

\beq
\sigma_{t}^2=\sigma_{tot}^2-\sigma_{r}^2=(\sigma_x^2+\sigma_y^2+\sigma_z^2)-\sigma_{r}^2 .  \label{sigma}
\eeq

\noindent
While $\beta$ and $\gamma$ can be directly computed in the FIE and RIE cases, in order to calculate them in the LIE scenario we first need to derive the line-of-sight component of the velocity vectors of the subhaloes. The line-of-sight velocity depends on the viewing angle of the host which is unknown in our simulations. We thus randomly rotate each host and its subhaloes  $N_{rot}=5$ times, taking the mean of all these resulting line-of-sight velocity to compute $\beta$. We perform a small number of rotations of the whole system since otherwise, by averaging over a higher number of rotations, we converge to the FIE case. The same methodology has been applied to the PIE case where we additionally had to project the distances of the tracers objects into the observers plane in order to obtain the $\gamma$ parameter.\\

\subsubsection{The host halo parameter: $\alpha$} \label{sec:alpha_param}

To get the value of $\alpha$, we must recall that since our haloes are not scale-free but rather follow a NFW profile \citep{Navarro96}, the applicability of a power-law is limited. While for a pure scale-free model it is irrelevant whether we fit the gravitational potential, the density or the mass profile of the host halo (see W10), it will most certainly lead to differences when the scale-invariance is broken. Recall that for a scale-free model:

\beq
\phi(r) \propto r^{-\alpha} \ \ \Leftrightarrow \ \ \rho(r)\propto r^{-\alpha-2} \ \ \Leftrightarrow \ \ M(r)\propto r^{1-\alpha} .
\label{from_potential_to_mass}
\eeq

\noindent
For a NFW object however, we must identify which quantity is the most suitable to be fitted, and we decided to use the cumulative mass profile $M(r)$ since this is the least noisy from a numerical point of view. 

\begin{table}
 \caption{Value of the $\alpha$ parameter and its fractional error $\Delta\alpha/\alpha$ obtained by fitting the numerical mass profiles of the MW and M31 over different radial ranges.}
\begin{center}
\begin{tabular}{lllll}
\hline
\hline
&$MW$&&$M31$&\\
\hline
\hline
Radial Range & $\alpha$  & $ \frac{\Delta\alpha}{\alpha}$& $\alpha$ &  $  \frac{\Delta\alpha}{\alpha}$   \\
\hline
\hline
$[0,1]R_{\rm vir}$                & -0.034 & 70 \%& -0.052& 60\%  \\
$[0.4,1]R_{\rm vir} $            &  0.302 & 8\%    &  0.266&  9\%\\
$[0.8,1]R_{\rm vir}  $           &  0.398 & 3\%    &  0.402 & 7\% \\	

 \hline
 \hline

\end{tabular}
\end{center}
\label{tab:alpha}
\end{table}

Furthermore, since our halo does not follow a scale-free profile (either in mass or in potential), the actual value of $\alpha$ depends on the radial range used to fit it, i.e. $\alpha(r)\neq const$.\\ 
We thus provide, in \Tab{tab:alpha}, the numerically fitted values of $\alpha$, obtained by fitting the total mass halo profile in different radial ranges, specifically in the total range, in the outermost one, and in the intermediate range, together with their fractional relative errors, where we indicate with $\Delta\alpha$ the $1\sigma$ error on $\alpha$ as found from the fitting routine.

 As in the previous case of the $\gamma$ parameter, we used a Poissonian weight ($1/N_{sub}$) to associate errors to the data during the fit: as expected, the smallest relative error is obtained in the outermost radial range, confirming that in this regime the host density profile is best approximated as being of scale-free nature.

We obtained for the MW and M31 in the total radial range a value of $\alpha=-0.034$ and $\alpha=-0.052$ respectively, as listed already in \Tab{tab:MWM31}, while we can observe how the $\alpha$ value increases when we move to the outer part of the halo, as expected if the halo is following a NFW profile, since it gets steeper towards the outer part of the distribution.

While using the numerical mass profile given by the simulation data is actually a self-consistent way to obtain $\alpha$, we note that an observer would require a mass model to actually determine the $\alpha$ parameter to be used with the mass estimators. Since an observer does not have any a-priori knowledge of the radial mass distribution (or potential) of the host halo, an analytical  profile must be assumed. Note that W10 showed that for an object following a NFW profile the typical value of $\alpha$ is $\approx0.55$, based upon fitting a NFW potential in the range $[10,300]$~kpc to a power-law $\phi\propto r^{-\alpha}$ and assuming to have hosts with concentration between $c=18$ and $c=8$. Given the uncertainty on the actual density profile of the real hosts, we decided to allow for the estimates of $\alpha$ in two different ways:

\begin{enumerate}
\item using the values derived by fitting our numerical profile at different radial ranges, or
\item using the relation $\alpha = \gamma -2$, which holds true if the subhaloes are tracking the total gravitating mass of the hosts.
\end{enumerate}

\subsection{Results for the simulated MW and M31} \label{sec:results}

The application of the scale-free mass estimator to the (observationally) unrealistic scenario in which we have $N\sim1000$ tracers, as found in our simulated haloes, gives excellent results for all the estimators FIE, RIE, LIE and PIE.
Using the error formulae listed in \Sec{sec:massdependance}, and allowing a maximum error on the calculation of the parameters $\alpha$, $\beta$ and $\gamma$ of about $\sim 20\%$, we obtained the MW mass at the $R_{\rm vir}=309$~kpc within a $5\%$ of uncertainty and the M31 mass at the $R_{\rm vir}=340$~kpc within a $3\%$ of error, respectively (FIE estimator). However, we decided not to show these results and rather focus on more interesting and practical situations where the number of tracer objects is limited and agrees better with the actual observed Local Group. We must note however, that part of our initial questions has been already answered by this exercise: the scale-free mass estimators are even applicable to a \textit{system} of host haloes such as the (observed) Local Group for which they were originally designed.

\subsubsection{Matching the number of the observed satellites} \label{sec:n_30}

As shown in \Tab{tab:MWM31}, the total number of subhaloes found within $300$~kpc in our simulations substantially differs from the number of observed satellites galaxies of the Milky Way and Andromeda within the same radius (the well known missing satellites problem, first addressed in \citet{Klypin99s} and \citet{Moore99}). Thus, we would like to calculate the accuracy of the mass estimators when the number of tracers is comparable to the real one, i.e. $N\sim30$ (we explicitly chose this number to be able to have a direct comparison with the W10 results, see for example their Fig.1). Further, the real case scenario is the one for which we have the radial velocities of the MW satellites and the line-of-sight velocities of the M31 tracers:  in the forthcoming analysis we will thus only use the RIE estimator for the Milky Way and the LIE one for the Andromeda galaxy.

From the total set of subhaloes we randomly selected $N=30$ objects that covered the total range within $r_{\rm out}<300$~kpc and computed their velocity anisotropy and their radial distribution, thus obtaining the $\beta$ and $\gamma$ coefficients. For this particular exercise, the $\alpha$ parameter was numerically evaluated using the three different radial ranges of the host mass profile listed in \Tab{tab:alpha} (ignoring the option to evaluate it as $\alpha=\gamma-2$ for the moment). For each of these values of $\alpha$ we performed 1000 random realization, we applied the scale-free mass estimator and we calculated the distribution of the ratio of the estimated over the actual mass, i.e. $M_{\rm est}/M_{\rm true}$; those distributions have then been fitted by a Gaussian curve eventually leading to the best-fit parameter $\mu$ and its standard deviation $\sigma$.

\begin{figure*}\begin{center}
$\begin{array}{cc}  \includegraphics[width=3in]{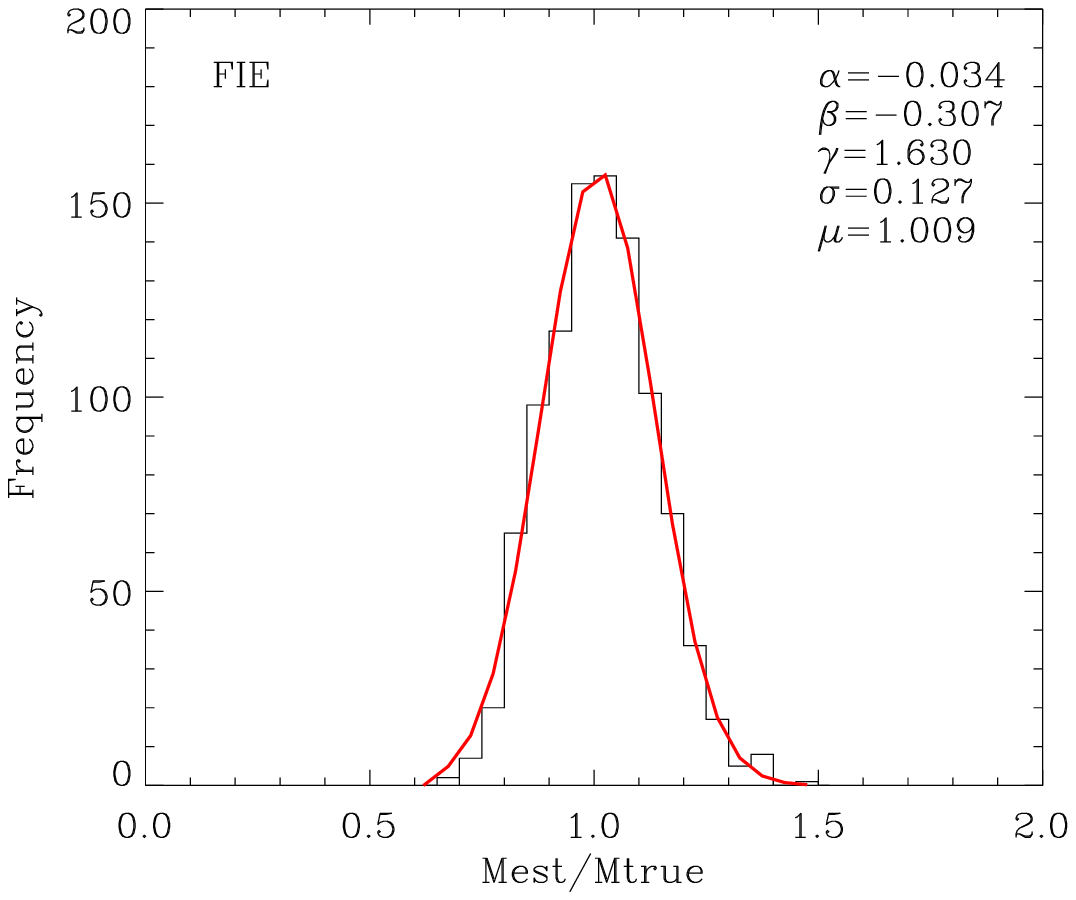}\includegraphics[width=3in]{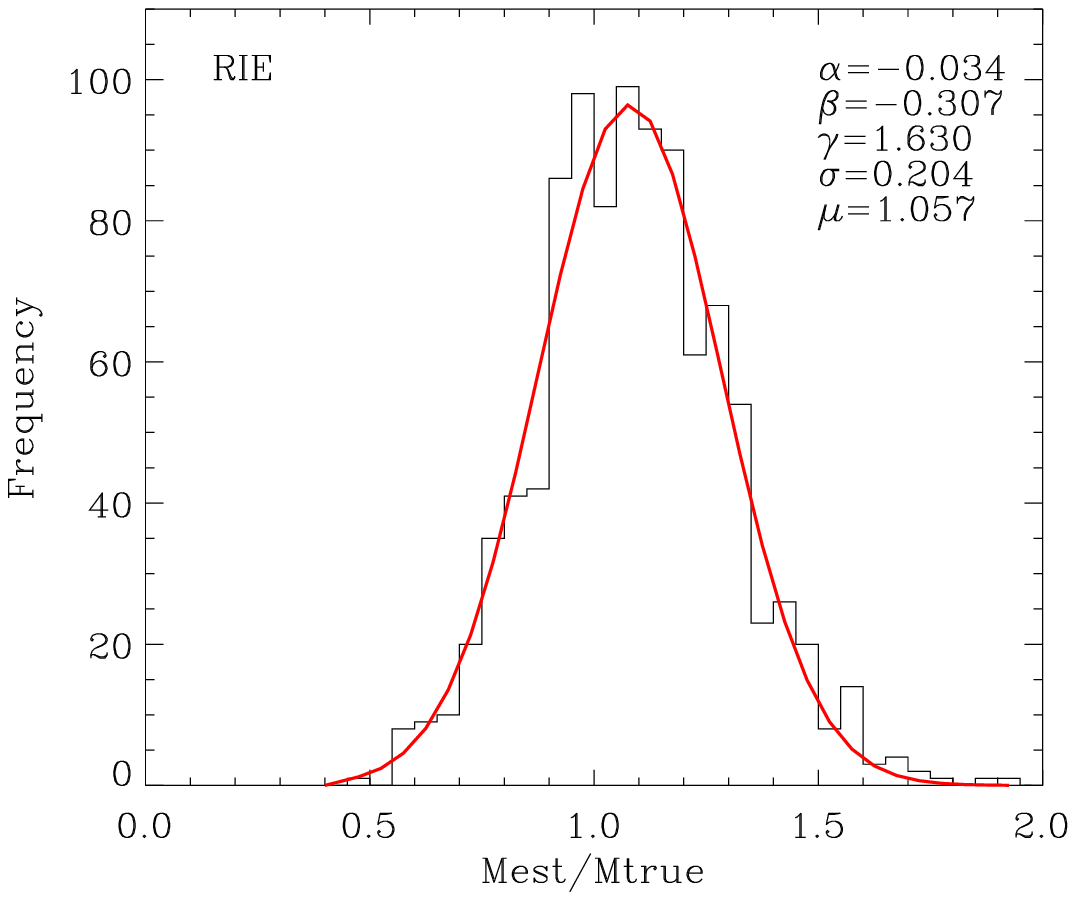}\end{array}$
  \caption{Distribution of 1000 realizations of the estimated mass over the real one for 30 subhaloes of the simulated Milky Way. The average value of the parameters $\beta$ and $\gamma$ obtained in each realization is shown. The best-fit Gaussian is also plotted, and its mean $\mu$ and standard deviation $\sigma$ are indicated. The left panel corresponds to the FIE estimator, the right panel to the RIE one.}
\label{fig:MW_30}
\end{center}
\end{figure*}

\begin{figure*}\begin{center}
$\begin{array}{cc}  \includegraphics[width=3in]{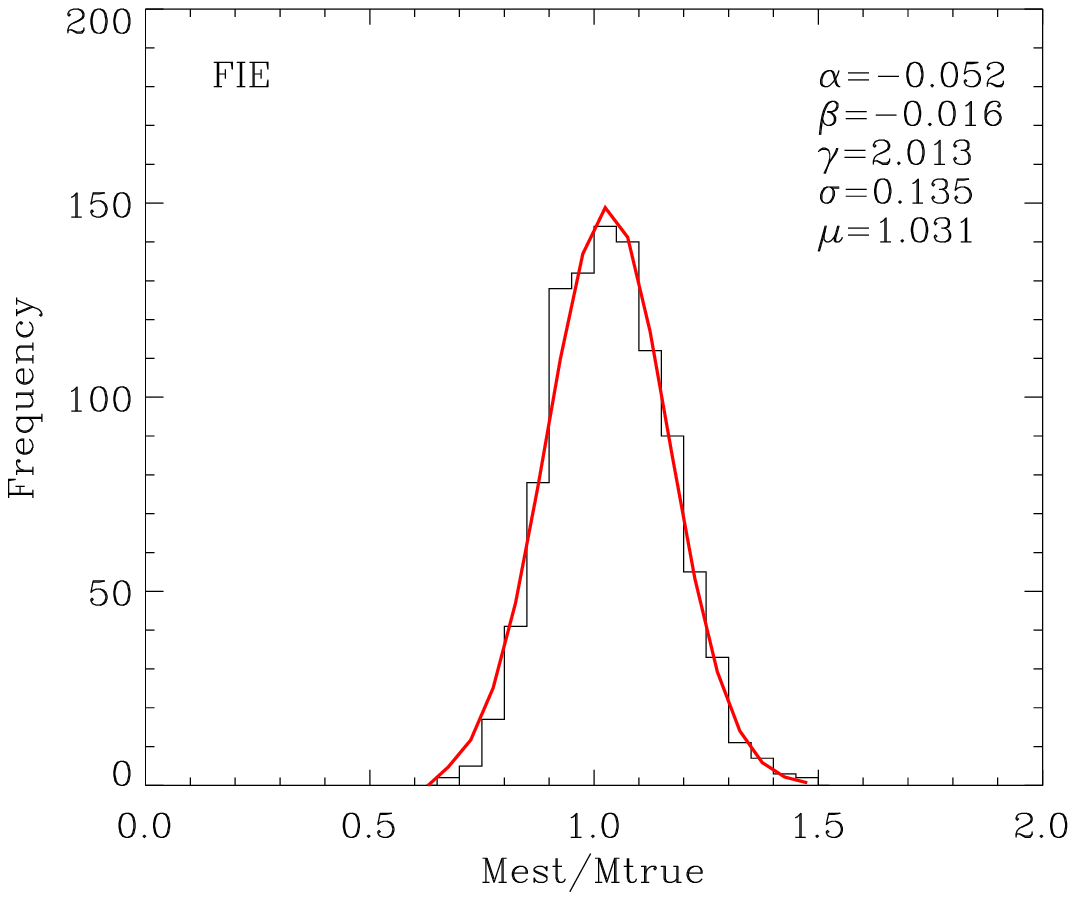}\includegraphics[width=3in]{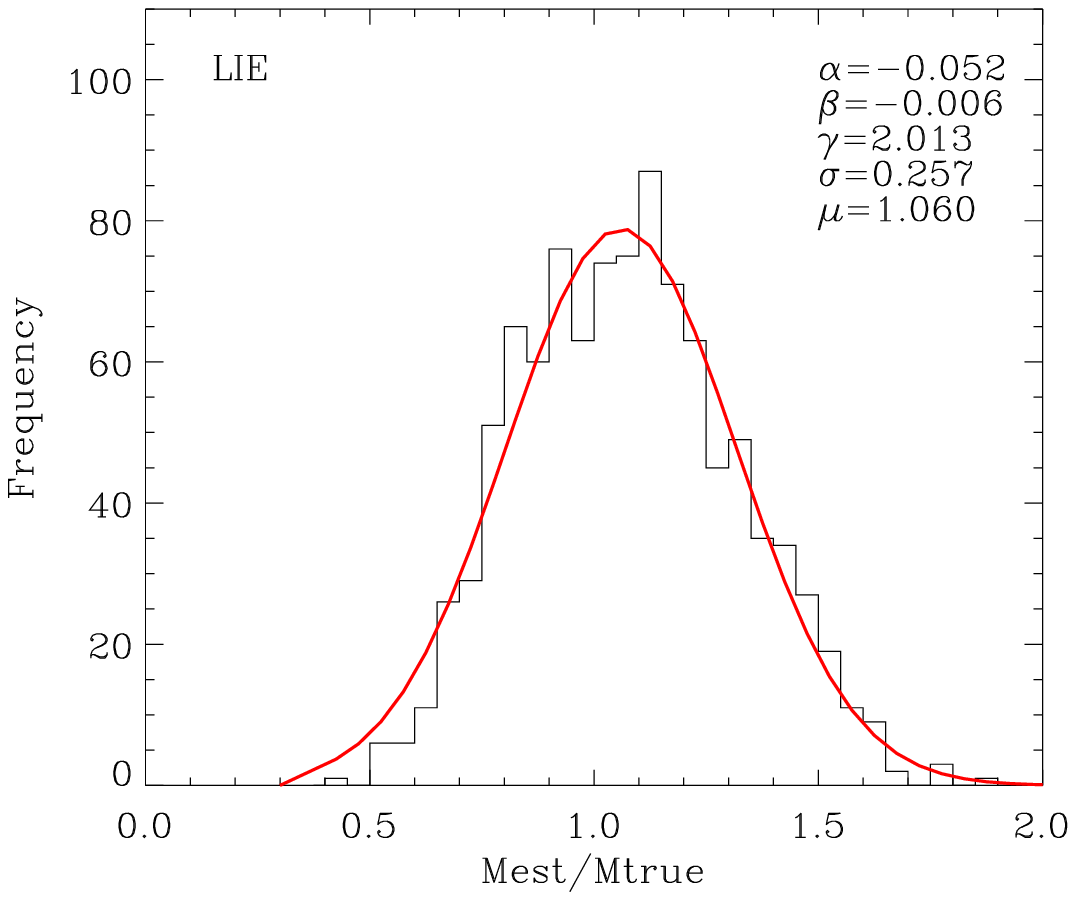}\end{array}$
  \caption{Distribution of 1000 realizations of the estimated mass over the real one for 30 subhaloes of the simulated Andromeda galaxy. The average value of the parameters $\beta$ and $\gamma$ obtained in each realization is shown. The best-fit Gaussian is also plotted, and its mean $\mu$ and standard deviation $\sigma$ are indicated. The left panel corresponds to the FIE estimator, the right panel to the LIE one.}
\label{fig:M31_30}
\end{center}
\end{figure*}

The results of these tests (for the $\alpha$ value evaluated from the total radial range, i.e. first line of \Tab{tab:alpha}) are summarized in \Fig{fig:MW_30} for the MW and \Fig{fig:M31_30} for M31 where we plot in the left panels the distributions of $M_{\rm est}/M_{\rm true}$ for the FIE mass estimators and in the right panels the RIE (MW) and LIE (M31), respectively. The legends of each panel further list the three parameters $\alpha$, $\gamma$, and $\beta$ relevant for the respective mass estimator (where $\beta$ and $\gamma$ represent the average value over the total 1000 realizations) alongside the peak and standard deviation of the best fit Gaussian. Note that the standard deviation is compatible with $1/\sqrt {N_{sub}}$ where $N_{sub}$ is the number of used tracers, and it increases when only radial velocities (or line-of-sight ones) are used. Remarkably, the mean of the distribution stays always very close to $\mu=1.0$: the mass estimators are thus unbiased with respect to the number of used objects. We repeated the above mentioned analysis for the other values of $\alpha$ listed in \Tab{tab:alpha}, and found practically indistinguishable results: the best-fit $\mu$ and $\sigma$ values are given in  in \Tab{tab:mu_MW}.

In summary, we found that for both host systems the mass is always recovered within a few percent of error when restricting the analysis to 30 randomly selected subhaloes each. 

\begin{table}
 \caption{Mean value and standard deviation of the Gaussian distribution of the estimated mass over the true mass for the MW and M31 hosts, using the RIE and the LIE, respectively, for the three considered values of $\alpha$ (cf. \Tab{tab:alpha}). The number of subhaloes has been limited to 30 random ones, and they have been used to compute $\beta$ and $\gamma$ for each realization. The average values of these parameters over the total $N=1000$ realizations are $\gamma=1.63\pm0.12$ and $\beta=-0.307\pm0.061$ for the MW, and $\gamma=2.013\pm0.013$ and $\beta=-0.006\pm0.001$ for M31.}
\begin{center}
\begin{tabular}{||llllllll}
\hline
\hline
&$MW$&&&$M31$&\\
\hline
\hline
 $\alpha$&$\mu$&$\sigma$&$\alpha$&$\mu$&$\sigma$ \\
\hline
\hline
 -0.034& 1.057 & 0.204  &-0.052&1.060 & 0.257\\
0.302 &  0.992 & 0.177  &0.266&  0.958 &0.167\\
0.398 &  0.975 & 0.169  &0.402&  0.931 &0.159\\	

 \hline
 \hline

\end{tabular}
\end{center}
\label{tab:mu_MW}
\end{table}

\subsubsection{Matching the radial number distribution of the observed satellites} \label{sec:gamma_2,8}
\begin{figure}\begin{center}
\includegraphics[width=3in]{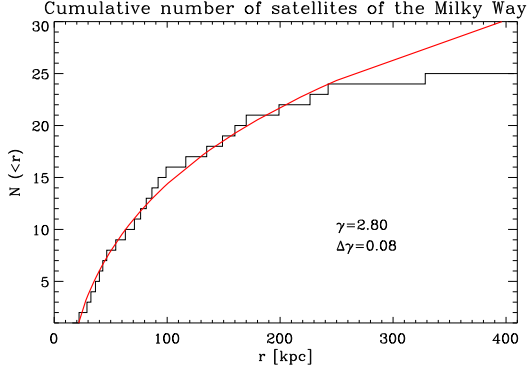}
  \caption{Radial distribution of the observed MW satellites and corresponding best fit in the range $r< 300$~kpc.}
\label{fig:MW_power_law}
\end{center}
\end{figure}

While using 30 randomly chosen subhaloes leads to exceedingly well recovered host masses, we acknowledge that our model subhaloes (for the MW) do not follow the same radial distribution as the observed ones (why this is the case is substance for yet another paper and shall not be addressed here): we list in \Tab{tab:satellites} the distances to all presently known MW satellites \citep[taken from][]{Wadepuhl10} alongside their masses and use this data to obtain the observed $\gamma\pm\Delta\gamma=2.80\pm0.08$ by fitting the radial distribution to a power-law in \Fig{fig:MW_power_law}. Please note that we only focus on the MW's subhaloes here, as in the case of M31 the $\gamma=2.013$ coefficient is very similar to the one obtained from the observed satellites distribution (see W10).

From the total set of subhaloes in our numerical MW, we constructed a subset of 30 tracers by selecting those objects that follow the radial distribution $N(<r)\propto r^{3-2.8}$. Further care was taken to verify that the randomly selected subhaloes always cover the (observational) radial range up to $\sim300$kpc. While the $\gamma=2.8$ is fixed by construction the $\beta$ has always been derived from this subset; for the $\alpha$ we first used, again, the three values listed in \Tab{tab:alpha}, and we found a notable bias in the Gaussian distribution of $M_{\rm est}/M_{\rm true}$, as high as the $80\%$: this choice of $\alpha$ does not provide the expected host mass. Thus, we secondly decided to verify if the assumption that the tracers are tracking the total gravitating mass of the host can provide a better constrain on the value of $\alpha$, i.e. using the relation $\alpha = \gamma-2$. In this case, without making any fits to the numerical shape of the host profile, we actually found results in excellent agreement with the true mass, as shown in the left panel of \Fig{fig:MW_30_gamma2,8}. In the right panel of the same figure we show the distribution obtained when yet another additional constraint was added, i.e. we selected only those subhaloes whose mass lies between $5\cdot10^6<M/M_{\odot}<1\cdot10^8$, in order to resemble the average mass of the observed MW satellites (see \Tab{tab:satellites}). Also in this case we can observe that the Gaussian is peaked very close to 1.0, at $\mu=1.016$.

We finally decided to also test and use the suggested value of $\alpha=0.55$ (W10), but we actually obtained a Gaussian mean value for $\mu$ that is biased by approximately $30\%$ towards large estimated masses.

In summary, even when restricting the subhaloes to follow the same power-law as the observed satellites within the same mass range, the scale-free mass estimators are capable of recovering the true mass of our constrained MW and M31 if one chooses to use $\alpha=\gamma-2$ (being close to the isotropic regime, i.e. $\beta=0$ and as far out as $\beta=-0.5$).

\begin{figure*}\begin{center}
$\begin{array}{cc}  \includegraphics[width=3in]{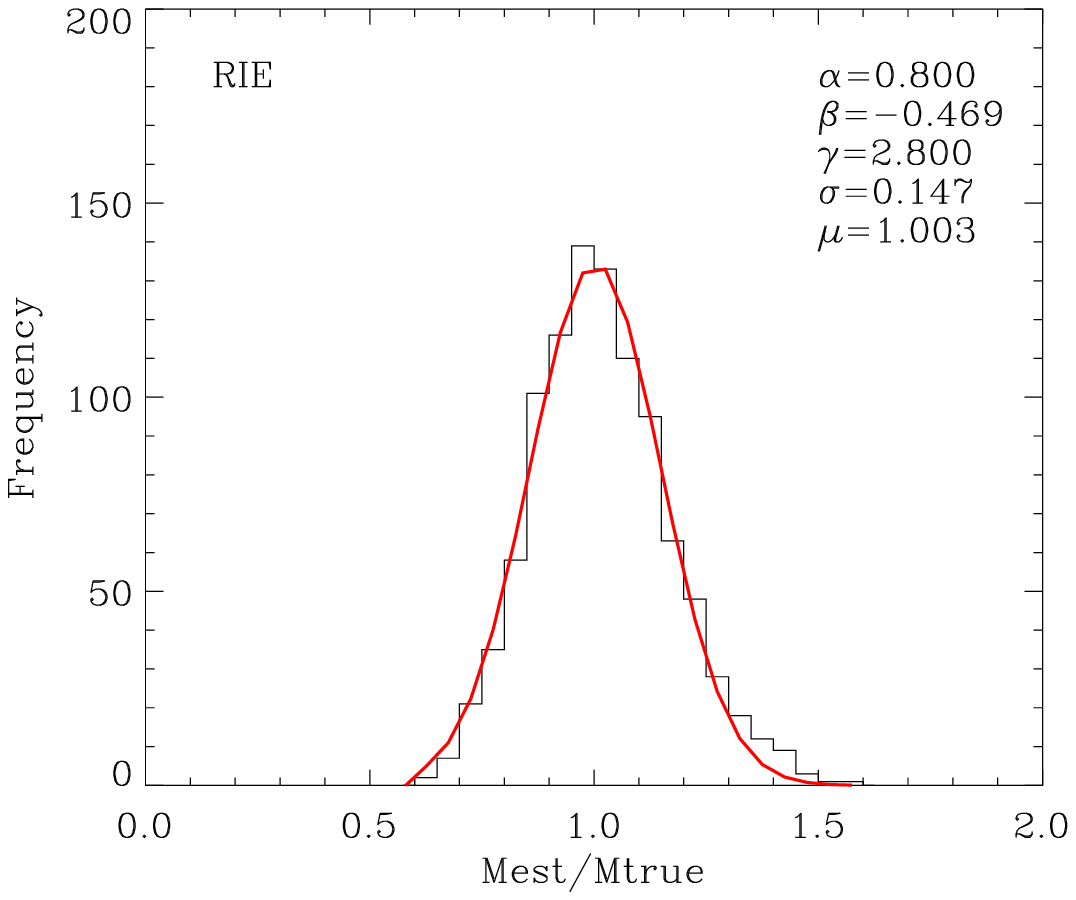}\includegraphics[width=3in]{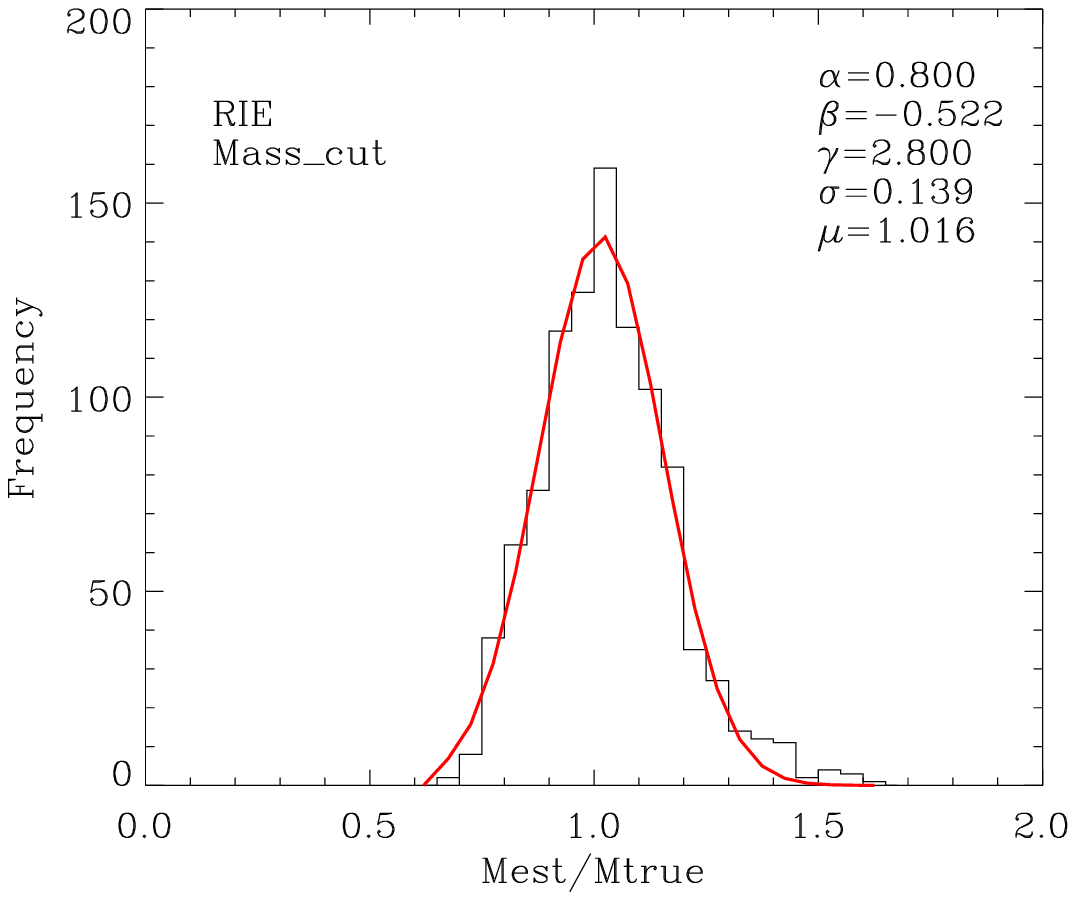}\end{array}$
  \caption{Distribution of the estimated RIE mass over the real one for 30 subhaloes of the Milky Way and 1000 realizations. The subhaloes have been selected following the power law with $\gamma=2.8$ (left panel) as well as additionally also constraining them to lie within the observed mass range (right panel). Note that in both cases $\alpha$ has been determined as $\alpha=\gamma-2$.}
\label{fig:MW_30_gamma2,8}
\end{center}
\end{figure*}

\begin{table}
 \caption{List of the MW satellites used in this work, corresponding to those lying within 300kpc from the galactic center and with measured line-of-sight velocities. The Galactocentric distances $D$ are in kpc. The values are from: (a) \citet{Martin08}, (b) \citet{Mateo98}, (c) \citet{Belokurov08}, (d) \citet{vandenBergh94}, (e) \citet{Belokurov09}, (f) \citet{Simon07}, (g) \citet{Bekki08}, (h) \citet{vandenBergh00}.}
\begin{center}
\begin{tabular}{lll}
\hline
\hline
Name & $D[kpc]$  &  $Mass[10^6M_{\odot}]$ \\
\hline
\hline
$Boo I^{(a)}$            & $66\pm3$ &    -                                             \\
$Boo II^{(a)}          $           &  $42\pm8$ &   -					\\
$Carina^{(b)}         $        & $ 101\pm5$& $13$					\\	
$Com^{(a)}              $       &  $44\pm4$ &$1.2\pm0.4^{(f)}$				\\
$CVn I^{(a)}              $     &$218\pm10$&$27\pm4^{(f)}$				\\	
$CVn II^{(a)}              $    &$160^{+4}_{-5}$ &$2.4\pm1.1^{(f)}$			\\
$Draco^{(a)}                $  &    $76\pm5$ &$22$           				\\
$Fornax^{(b)}                 $ &    $138\pm8$ & $68$    				\\	
$Her ^{(a)}                       $&$132\pm12$ &$7.1\pm2.6^{(f)}$			\\	
$Leo I^{(b)}            $           &  $250\pm30$ &$22$				\\
$Leo II^{(b)}             $        & $205\pm12$&$9.7$					\\
$Leo IV^{(a)}             $      &$160^{+15}_{-14}$ &$1.4\pm1.5^{(f)}$		\\
$Leo V^{(c)}                 $     &$180$& -							\\
$LMC^{(d)}                    $ &$49$ & $10.000^{(g)}$					\\	
$Sag^{(b)}                  $       &$24\pm2$& $150^{(h)}$						\\
$Sculptor^{(b)}             $    & $79\pm4$&$6.4$						\\
$Seg I^{(a)}                    $   &$23\pm2$&-							\\
$Seg II^{(e)}                     $&$35$ & $0.55^{+1.1}_{-0.3}$			\\			
$Sextans^{(b)}       $          &$86\pm4$ &$19$					\\
$SMC^{(d)}              $       &$58$& $400^{(g)}$						\\
$UMa I^{(a)}               $     &$96.8\pm4$ & $15\pm4^{(f)}$				\\
$UMa II^{(a)}                $   &$30\pm5$ & $4.9\pm2.2^{(f)}$				\\
$UMi^{(b)}                        $& $66\pm3$ & $23$						\\	
$Wil 1^{(a)}                        $& $38\pm7$ & -					\\

 \hline
 \hline

\end{tabular}
\end{center}
\label{tab:satellites}
\end{table}

\subsubsection{Do we require a host mass profile or simply $\alpha=\gamma-2$?} \label{sec:alpha=gamma-2}
The analysis in the previous subsection has shown that simply using $\alpha=\gamma-2$ actually leads to excellent results for the scale-free mass estimators when applied to our constrained Local Group and a subhalo population restricted to follow the observed one as closely as possible. But can this finding be generalized, at least with respects to our  simulation?

\begin{figure*}\begin{center}
$\begin{array}{cc}  \includegraphics[width=3in]{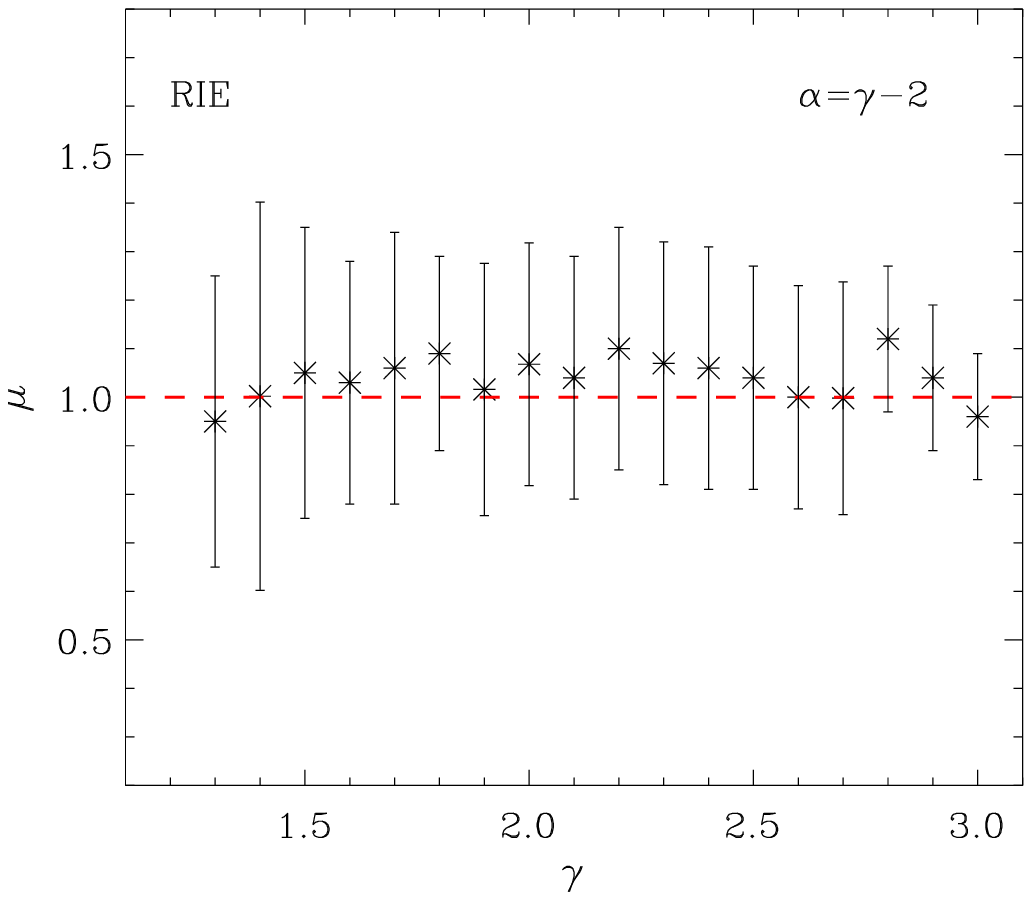}\includegraphics[width=3in]{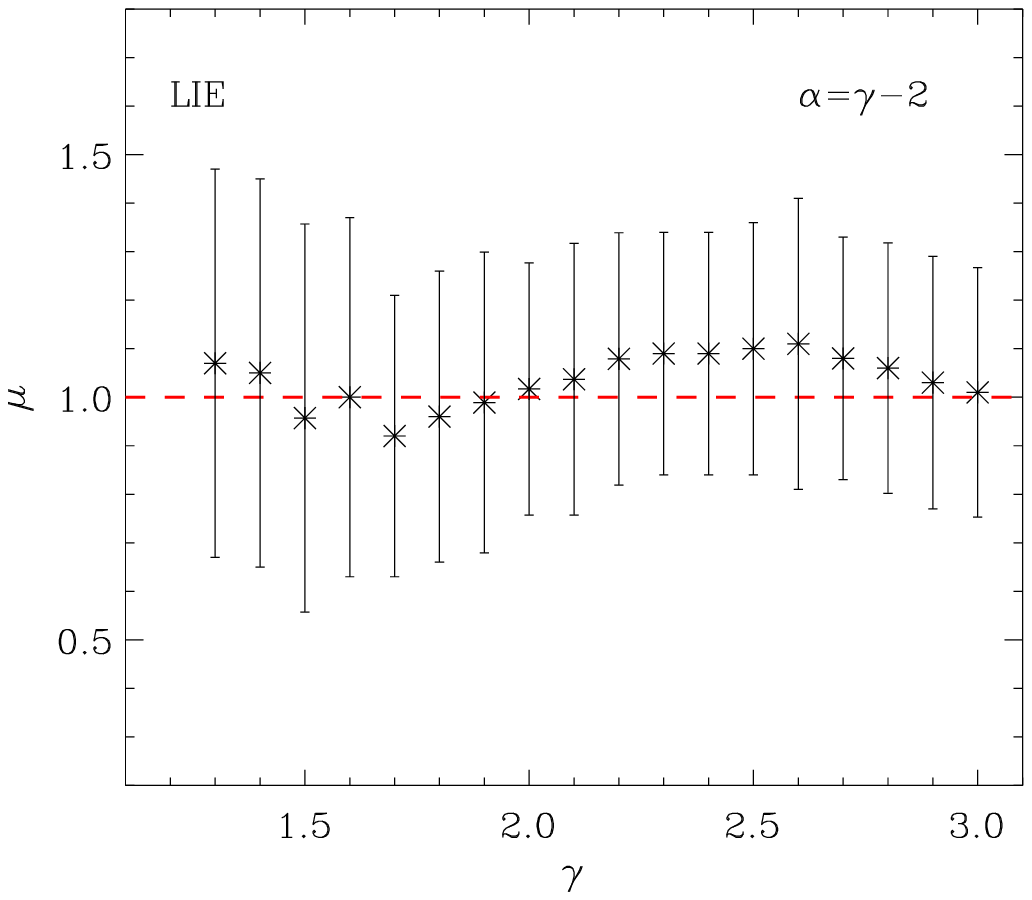}\end{array}$
  \caption{Mean value $\mu$ of the best fit Gaussian distribution for different values of $\gamma$ and correspondingly $\alpha=\gamma-2$. The anisotropy parameter lies between $-0.3<\beta<0.1$. The distribution is based upon 1000 realization. MW, using the RIE, left panel, M31, using the LIE, right panel.}
\label{fig:mu_versus_gamma}
\end{center}
\end{figure*}

To verify that the assumption $\alpha = \gamma -2$ holds true in general, we select the subhaloes of both MW and M31 in order to follow different radial distributions, according to $N(<r)\propto r^{3-\gamma}$, where we allowed the $\gamma$ to vary between $1.5$ and $3.0$. The $\alpha$ coefficient was then calculated consequently, while the $\beta$, as usual, came from the selected satellites velocity dispersions. In this way we selected $N=30$ subhaloes again for 1000 times and we obtained Gaussian distributions of the $M_{\rm est}/M_{\rm true}$ quantity. We show the best-fit $\mu$, with the standard deviation $\sigma$ as error bars, as a function of $\gamma$ in \Fig{fig:mu_versus_gamma} for the MW (left panel, RIE only) and M31 (right panel, LIE only). The anisotropy parameter was always found to be close to $\beta \sim0$, with a maximum variation between $-0.3<\beta<0.1$, indicating that we are in the isotropic regime. We would like to highlight that despite other choices of $\alpha$ may in principle be possible, as demonstrated in \Sec{sec:n_30}, the simple assumption $\alpha=\gamma-2$ provides always the best estimation for the host mass, whose associated Gaussian distribution has mean value $\mu$ closer to 1 and smaller standard deviation $\sigma$.

We see that the simple assumption, that the satellites track the total mass of the host halo, is sufficient to give an excellent estimation of the parameter $\alpha$ to be used. The suggested value of $\alpha=0.55$, indicative of a NFW halo potential, is thus recommendable when the satellite distribution follows a power law with exponent $\gamma\sim 2.5$: these values have been already successfully used in \citet{Deason11} and \citet{Evans11}. This is of fundamental importance for observations: without having any a-priori knowledge about the host halo density profile, we can simply use the value $\alpha= \gamma-2$ once we have calculated $\gamma$ from the  satellite distribution. This condition has been verified in our constrained simulations only, in which the anisotropy parameter is always $\beta \sim 0$: care should be taken when dealing with satellite populations whose $\beta$ is highly anisotropic.

In summary, we have shown that our two model hosts within the simulated constrained Local Group allow the application of scale-free mass estimators to them. And, for as long as we are in the isotropic regime in which $\beta=0$, the simplifying assumption of $\alpha=\gamma-2$ can be used. This alleviates the need to derive this parameter from a model of the host potential or mass profile.

\subsection{Exploring the influence of the MW and M31 on each other} \label{sec:exploring}
\subsubsection{Subhaloes (not) facing the opposite host} \label{sec:facing}

The fact that the MW and M31 hosts are close to each other, forming a binary galactic system, poses the question if the mass estimators work even in the situation in which we only consider satellites in between the two haloes. To shed light on this issue we begin by separating the MW halo into two hemispheres, defined as ``facing'' and ``non-facing'' M31 (we perform the same test for the M31 halo, too). Subhaloes are then grouped by the hemisphere they sit in. We remark that this is a purely spatial cut, to investigate if the proximity of the companion host causes some bias on the mass estimator. 

The facing/non-facing subhaloes of the MW are selected according to a radial number distribution $N(<r)\propto r^{3-\gamma}$ with $\gamma=1.63$, and consequently $\alpha=\gamma-2=-0.37$, while the M31 subhaloes are selected in order to follow the distribution with $\gamma=2.013$ and $\alpha=0.013$. We chose these value to match the parameters found in \Sec{sec:n_30}, but verified that our results are not affected by the choice of the specific power law, as already expected (cf. \Fig{fig:mu_versus_gamma}). Finally, we randomly selected $N=30$ subhaloes from each of the two facing/non-facing subsets, repeating the analysis 1000 times and computing each time the parameter $\beta$ and the estimated mass. The anisotropy parameter, for both MW and M31 and in every subset of objects considered, has been found to be very close to $0$ again, lying between $-0.35<\beta<0.05$. We are hence in a situation to explore the influence of the two hosts on each other: with the $\beta$ close to $0$ and the subhaloes selected to follow a fixed power law (thus without errors associated to the 3 main parameters) we can affirm that any deviation in the mass estimation should now be attributed to the subhaloes facing or not the other host.

In \Tab{tab:facing} we show the results of the mean value and standard deviation of the Gaussian distribution of the estimated mass over the true one for the MW (by applying the RIE) and for M31 (using the LIE). 
We show the $\mu$ obtained by using the facing subhaloes, the non-facing subhaloes and also the total set. We did this exercise for two different values for $r_{\rm out}$, thus computing the host mass up to this outer radius: in this way we should be able to observe if the proximity or, on the contrary, the distance of the subhaloes population to the opposite host has some influence as well. For the MW host we can observe that the estimator performs equally well when using the facing or non-facing objects, for each of the two $r_{\rm out}$ values used. In the case of M31, instead, the non-facing population of subhaloes seems to give better results in the estimation of the host mass, while the facing objects lead to a Gaussian distribution whose mean value is slightly biased ($\mu=1.11$) when we considered the $r_{\rm out}=221$~kpc. Given the high uncertainties associated to this biased result however, we can conclude that each of the main galaxies  does not influence the subhalo population of the other one.

\begin{table*}
 \caption{Milky Way and Andromeda galaxy mass estimation (RIE for MW, LIE for M31) using the subset of subhaloes facing and non-facing the companion host. The mean value $\mu$ and standard deviation $\sigma$ of the Gaussian distribution of the estimated mass over the true one are shown, obtained by selecting N=30 subhaloes from the facing or not facing subsets, repeating 1000 realizations and evaluating the mass at two different $r_{\rm out}$.}
\begin{center}
\begin{tabular}{llll}
\hline
\hline
Host $r_{\rm out}$  & All Subs & Facing Subs  &  Not Facing Subs  \\
\hline
$r_{\rm out,MW}$&$\mu\pm\sigma$&$\mu\pm\sigma$&$\mu\pm\sigma$\\
\hline 
$309$~kpc&                                $0.97\pm0.12$ & $1.05\pm0.16$ &$0.85\pm0.24$ \\
$197$~kpc&                                $0.98\pm0.12$ & $1.01\pm0.12$ &$1.03\pm0.26$ \\
                                                             
\hline
 \hline

$r_{\rm out,M31}$&$\mu\pm\sigma$&$\mu\pm\sigma$&$\mu\pm\sigma$\\
\hline 
$340$~kpc&                                $1.02\pm0.19$ & $1.08\pm0.19$ &$0.92\pm0.15$ \\
$221$~kpc&                                $1.05\pm0.19$ & $1.11\pm0.22$ &$0.98\pm0.16$ \\
                                                             
\hline
 \hline
\end{tabular}
\end{center}
\label{tab:facing}
\end{table*}

\subsubsection{Renegade subhaloes} \label{sec:renegade}
As already discussed, we call renegade subhaloes those objects, in our simulations, that change their affiliation from one of the two prominent hosts in
the Local Group to the other \citep{Knebe11b}. We were able to identify 129 renegade objects, 57 of which belonging to M31 at z=0, and the remaining 72 belonging to the MW. 

We thus examine the effect that this population of renegade subhaloes may have on the mass estimators: while in the previous studies we used the full set of subhaloes, automatically including also the renegade ones, we now want to restrict the analysis to the renegade subhaloes only in order to estimate the mass of MW and M31.

For each host we computed the anisotropy parameter and radial distribution of their respective renegade satellites, and found $\beta_{M31}=0.86$ - $\beta_{MW}=0.72$ and $\gamma_{M31}=2.15\pm0.21$ - $\gamma_{MW}=2.01\pm0.20$. The choice of the host parameter $\alpha$ is made considering its value in the total $[0,1]R_{\rm vir}$ range, or using the relation $\alpha=\gamma-2$ that we provided in the previous sections.
The resulting estimated masses are shown in \Tab{tab:renegade_tab}, in which we have used the FIE estimator in order to compare the effects of these renegade subhaloes in the same way for both hosts. This time, because of the small number of objects considered, we do not perform multiple realizations of the mass estimation, but only one: the errors associated with the mass are thus computed through the error propagation formula based on \Eq{mass_relative_variation} - \Eq{mass_relative_variation_PIE}, in which we further assume that $\beta$ is fixed, $\Delta\gamma/\gamma\sim10\%$ as obtained by the fitting routine and $\Delta\alpha/\alpha$ is listed in \Tab{tab:alpha}. We see that in the case of using the value of $\alpha$ from the total radial range, the mass estimator is biased for both hosts, with a large associated error. When using the relation $\alpha=\gamma-2$, instead, the mass of both Milky Way and Andromeda in recovered within a much smaller uncertainty. Is interesting to note how the relation between the host parameter $\alpha$ and the subhaloes distribution parameter $\gamma$ seems to hold true also in this case, in which the anisotropy parameter $\beta$ is substantially far from being isotropic. However the lack of statistic in this case, having at our disposal just one realization of a small number of renegade objects, prevents us from generalizing the finding of \Sec{sec:alpha=gamma-2} to this highly anisotropic case. The fact that $\beta\sim1$ for the renegade subhaloes means that these objects are mainly moving on radial orbits with respect to their hosts.
We conclude that the computation of the host mass based upon a family of renegade subhaloes gives results in agreement with the expected ones and hence these mass estimators will not be biased in case that renegade objects also exist in reality.

\begin{table}
 \caption{Estimation of the MW and M31 mass based upon renegade subhaloes only, FIE case. The parameter used are $\beta=0.72$ and $\gamma=2.01$ for the MW, $\beta=0.86$ and $\gamma=2.15$ for M31. We remind that the value of $\alpha$ from the total radial range is $\alpha=-0.034$ for the MW and $\alpha=-0.052$ for M31, as in \Tab{tab:alpha}.} 
\begin{center}
\begin{tabular}{lll}
\hline
\hline
&MW&M31\\
\hline
\hline
$\alpha$& \multicolumn{2}{c}{$(M_{est}\pm \Delta M_{est})/M_{true}$}\\
\hline
\hline
$[0,1]R_{\rm vir}$& $0.92 \pm 0.33$ &  $0.75 \pm 0.40$ \\
$\gamma-2 $       &  $1.00 \pm 0.21$ & $0.97 \pm  0.16 $ \\

 \hline
 \hline

\end{tabular}
\end{center}
\label{tab:renegade_tab}
\end{table}

\subsubsection{Unbound subhaloes} \label{sec:unbound}
For all previous results we did not test whether or not a subhalo is gravitationally bound to its host; we simply used a spatial criterion to determine its affiliation as this is how satellites are often defined in the observations.
Now instead, we impose an additional constraint on our subhalo population: its velocity has to be lower than the local escape velocity $v_{esc}$ of the halo at that distance.
Following this criterion, we find that about the $3\%$ of the subhaloes inside the virial radius of each host are unbound. 
As expected, most of them lie near by the virial radius, where the $v_{esc}$ is lower and the effects of the proximity of the other host are more important. 
We thus quantify the effects that unbound subhaloes have on the mass estimators.
This is an interesting test, as it corresponds to the real case scenario in which the affiliation of a tracer object is not clear and could be erroneously included into the calculation of the mass of a host.

We repeat our previous methodology by evaluating the MW and M31 mass 1000 times with a subset of $N=30$ subhaloes, this time including one, two or three unbound subhaloes. In order to ensure an unbound subhalo is included, we explicitly substitute in each realization, one, two or three of the 30 subhaloes with an unbound one. We then computed the $\beta$ and $\gamma$ coefficients for each realization, and used the formula $\alpha=\gamma-2$.
We verify that the inclusion of a single unbound subhalo leads to mass estimators which are slightly biased towards larger masses: we obtained, for both M31 and MW, a Gaussian peaked at $\mu=1.04$ with $\sigma=0.14$.
When including two unbound subhaloes, we found an higher deviation, with $\mu=1.12$ and $\sigma=0.13$.
Finally, forcing three unbound subhaloes to be included in each 30 subhalo subsample, we obtain a Gaussian peaked at $\mu=1.17$ and $\sigma=0.14$.
We performed the calculation using the FIE estimator but verify that our results are the same in the RIE and LIE case, giving similar values for the mean of the distribution $\mu$, and increasing standard deviations $\sigma$ with respect to the FIE case.

We note that the results presented in the previous sections are not significantly affected by the presence of unbound subhaloes, as in that case the probability that in a given realization of $N=30$ randomly picked subhaloes one was unbound is $P\approx37\%$\footnote{This probability can be computed using the hypergeometric distribution, which, in our case and for the M31 and MW, respectively, describes the probability to get one unbound subhalo within $k=30$ randomly drawn objects from a total subhaloes population of size $N=1405(1205)$ in which the unbound objects are $n=45(36)$, thus the $\sim3\%$ of the total.
\beq
P=\binom {n} {1}  \binom {N-n} {k-1}/ \binom {N} {k}
\eeq}, due to the fact that unbound subshaloes make up just $3\%$ of the full subhalo population.
In this last test, instead, the probability that one object is unbound, over the $N=30$ subhaloes used in each realization, is  $P=100\%$, because we deliberately replaced one random subhalo with an unbound one. Thus, we expect that the error on the mean value of the Gaussian distributions in the previous analysis, caused by the possible inclusion of one unbound object, is $100/37\approx2.7$ times lower than the error made here, when one subhalo is forced to be unbound.
Looking at the $\mu=1.04$ obtained in this section, for the FIE case when we used a single unbound subhalo, we see that the $4\%$ of deviation from the expected value will be reduced of a factor $2.7$, giving negligible errors.
We are further reassured by having performed the analysis of \Sec{sec:results} also by explicitly neglecting the unbound objects, and we have observed no significant differences in the results already presented.

To conclude, in this section we demonstrate that, being sure of having included unbound subhaloes, this inclusion causes an overestimate of the host mass, in agreement with the results of \citet{Deason11}.
The more unbound objects we include into the mass estimator, the more biased the final mass is. Care should thus be taken when considering objects at the "edge" of a galaxy halo, as they may be not bound to it.

\section{Summary \& Conclusions} \label{sec:conclusions}
We verified the accuracy of the scale-free mass estimators recently proposed by \citet[][W10]{Watkins10} when applied to the two dominant Local Group host haloes, the MW and M31, by using Constrained Local UniversE Simulations (CLUES). These scale-free mass estimators assume that all the relevant informations about the enclosed mass of a halo are contained in the properties of its satellites (or any other tracer population), such as distances and velocities, which can both be given as either projected or full $3D$ data. The importance of such estimators resides in the fact that the full six-dimensional phase-space information of all celestial bodies down to the very faint magnitude $G\approx20$ mag will soon be available thanks to the upcoming GAIA mission.
What makes the usage of these mass estimators so appealing? After three years of operation the Sloan Digital Sky Survey II (SDSS-II\footnote{\texttt{http://www.sdss.org/}}) discovered eight new dwarf galaxies, seven of them orbiting our Galaxy. The SDSS, which covered more than a quarter of the sky, essentially doubled the known number of MW satellite galaxies, helping close the gap between the observed number of dwarf satellites and theoretical predictions.
During its projected five-year mission, GAIA will scan the entire $41253$ square degrees of the sky, obtaining astrometric parameters (angular position, proper motion, and parallax) for roughly one billion stars.
Recently, \citet{An11} further investigated the benefits that the use of all this new proper motion data will introduce in the application of mass estimators. It is thus imperative to develop and test against simulations the mass estimator based entirely upon tracer objects, such as satellite galaxies. This issue has already been partially addressed in \citet{Deason11} and \citet{Evans11}, using the GIMIC suite of simulations, from which they selected a set of galaxies that resemble the Milky Way.

In this work we tested the scale-free mass estimators against Constrained Simulation of the Local Group, in which observational data of the nearby Universe is used as constraints on the initial conditions. These constrained simulations provide us with a unique Local Group seated in the correct environment, as opposed to un-constrained cosmological simulations, to make a direct comparison between numerical results and observations: verifying the goodness of the W10 mass estimators in our simulated LG should therefore been seen as complementary to the already published work on their credibility and as an extension to a system resembling as closely as possible the real Local Group.

Our motivation is driven by the fact that the Local Group likely is a rather special (binary) system of galaxies featuring backsplash galaxies \citep{Knebe11a}, renegade satellites \citep{Knebe11b} and preferential infall: \citet{Libeskind10infall} have studied the simulated MW and M31 galaxies in the CLUES framework and have found a clear evidence for the anisotropic infall of subhaloes onto their respective hosts. This result has been recently corroborated by \citet{Keller11}, who examined the spatial distribution of the MW young halo globular clusters finding that they are anisotropically spatially distributed, sharing the same accreted origin as that of the MW's satellite galaxies. Our simulations also show the typical anisotropy in the distribution of subhaloes population, compatible with the observed classical MW satellites \citep[]{Kroupa05,Metz07,Metz08}, and as already found in other cosmological simulations \citep[cf.][]{Knebe04,Libeskind05,Zentner05}.  Furthermore, when comparing constrained against un-constrained simulations -- only 1-3\% of Local Group (candidates) share similar formation properties \citep{Forero11}. Thus, it is clear that our Local Group is a very special object in the Universe.

We first studied the sensitivity of the mass estimators with respect to their main parameters: $\alpha$, which describes the host halo scale-free gravity field, $\beta$, which corresponds to the satellites' velocity anisotropy, and $\gamma$, representing the exponent of the power law describing the number density of the tracer population. We found that for an external galaxy, such as M31, for which only line-of-sight informations are available, the greatest error comes from the uncertainty of $\alpha$ and $\gamma$, whereas the mass variations stemming from the anisotropy parameter $\beta$ are almost negligible in the interesting physical range.
On the other hand, the greatest concern in the estimation of the mass of our Galaxy comes from the $\beta$ parameter, as pointed out by W10: without precise information about the satellites' proper motions, the error introduced by using their radial velocities is sensibly high. Hopefully, future surveys (e.g. the GAIA mission) would be able to measure such proper motions, significantly improving the quality of these mass estimators.

We then applied the relevant mass estimators to the MW and M31 Local Group system as found in our constrained simulation. 
We found that all the estimators (FIE and RIE for the MW, FIE and LIE for M31) provide an unbiased results, with the mean of the $M_{est}/M_{true}$ distribution close to $\mu=1.0$ and its standard deviation scaling with $1/\sqrt{N_{sub}}$, even when a small ($N\sim30$) number of tracer objects are used, resembling the real case scenario of the known satellites galaxies. When selecting the subhaloes in order to follow a specific radial number distribution $N(<r)\propto r^{3-\gamma}$ with different $\gamma$, we found that, in the limit of the isotropic regime (i.e. $\beta\sim0$ and as far as $\beta=-0.5$ in our simulations) the assumption that the subhaloes are tracking the total mass of the host (i.e. $\alpha=\gamma-2$) is sufficient to get a very good estimate of the host mass.

We thus also investigated how the mass estimators work when using subhaloes that are closer or further away from the neighboring host, by restricting the analysis to the facing and non-facing hemispheres and calculating the mass at different values for $r_{\rm out}$: we observe that the two dominant hosts of the Local Group do not appear to influence its subhalo populations -- at least not when the applicability of the scale-free mass estimators is concerned.

Finally, we explored the possibility that using renegade subhaloes, i.e. subhaloes that change their affiliation from one of the two hosts to the other, in the estimation of the mass may cause a bias: we do not find evidence for this, on the contrary, we observe that the mass of both MW and M31 is recovered with a few percent of error when the assumption $\alpha=\gamma-2$ is made, even if the anisotropy parameter in this case is $\beta\sim0.7$, indicating that these objects are moving on radial orbits.

On the other hand, the inclusion of unbound objects, mainly found near the virial radius of each host, is able to cause an overestimate of the host mass, as high as the $20\%$ when 3 unbound subhaloes are used out of a total of 30 objects.
In this regard, care should be taken when dealing with tracer populations whose affiliation to each of the two host is not clear.
As long as boundness is verified however, the unique subhaloes population in our simulations, showing anisotropy in their spatial distribution, preferential infall \citep{Libeskind10infall} and even renegade objects \citep{Knebe11b}, does not prevent us from always recovering a good estimation of the host mass.

Hence, the most important finding of  this work is that satellite galaxies are well suited to ``weigh'' the MW's halo. Even with a small sample of just two dozen objects and despite anisotropic accretion, an anisotropic spatial distribution, different masses, sizes and histories, subhalo kinematics are dominated by the host potential, making satellite galaxies well suited for the problem at hand. We therefore conclude that the application of the scale-free mass estimators to either of the two dominant hosts of the Local Group provide credible results, it therefore appears safe to use it for the Local Group as already done by W10. 
To get a good estimation of a host mass, in the case of having the anisotropy parameter $\beta\sim0$, we recommend the use of the parameter $\alpha=\gamma-2$ where the $\gamma$ directly come from the observation of a satellite population.

In the future, sensitive surveys and space based telescopic missions will most likely improve both the census of satellite galaxies as well as our understanding of their proper motions: these new data will enhance the mass estimators making them more accurate than they are today.

\section*{Acknowledgements}
The authors thank the referee for the very detailed report and useful comments. This report led us to a number of substantial changes and additions which improved the paper very much. The simulations were performed and  analyzed at  the Leibniz Rechenzentrum Munich (LRZ) and at the Barcelona Supercomputing Center (BSC). We thank DEISA for giving us access to computing resources in these centers through the DECI projects SIMU-LU and SIMUGAL-LU.
YH has been partially supported by the Israel Science Foundation (13/08). NIL is supported by a grant by the Deutsche Forschungs Gemeinschaft. 
AK is supported by the {\it Spanish Ministerio de Ciencia e Innovaci\'on} (MICINN) in Spain through the Ramon y Cajal programme as well as the grants AYA 2009-13875-C03-02, AYA2009-12792-C03-03, CSD2009-00064, and CAM S2009/ESP-1496.
GY acknowledges support also from MICINN under research grants AYA2009-13875-C03-02, FPA2009-08958 and Consolider Ingenio SyeC CSD2007-0050.

\bibliographystyle{mn2e}
\bibliography{archive}

\bsp

\label{lastpage}

\end{document}